\documentclass[
 reprint,
superscriptaddress,
 amsmath,amssymb,
aps,
prd,
]{revtex4-2}

\usepackage{graphicx}
\usepackage{dcolumn}%
\usepackage{bm}%
\usepackage{xcolor}
\usepackage{physics}%
\usepackage{mathtools}%
\usepackage{hyperref}%
\usepackage{ulem}

\newcommand{\Xeff}{\chi_\mathrm{eff}}
\newcommand{\Xp}{\chi_\mathrm{p}}

\newcommand{\XIz}{\chi_{\mathrm{1z}}}

\newcommand{\XIIz}{\chi_{\mathrm{2z}}}

\newcommand{\XIp}{\chi_{\mathrm{1p}}}
\newcommand{\XIIp}{\chi_{\mathrm{2p}}}

\usepackage{color}

\hypersetup{
    colorlinks=true,
    linkcolor=blue,     
    citecolor=blue,     
    urlcolor=blue       
}

\begin{document}

\preprint{APS/123-QED}

\title{Impact of Spin Priors on the Population Inference of Merging Binary Black Holes
}

\author{Kazuya Kobayashi}
 \email{kazuya@icrr.u-tokyo.ac.jp}
\affiliation{
 Institute for Cosmic Ray Research, University of Tokyo, Kashiwanoha 5-1-5, Kashiwa, Chiba 277-8582 Japan
}

\author{Masaki Iwaya}
\affiliation{
 Gravity Exploration Institute, School of Physics and Astronomy, Cardiff University, 14-17 The Parade, Cardiff CF24 3AA, UK
}
\affiliation{
 Institute for Cosmic Ray Research, University of Tokyo, Kashiwanoha 5-1-5, Kashiwa, Chiba 277-8582 Japan
}

\author{Soichiro Morisaki}
\affiliation{
 Institute for Cosmic Ray Research, University of Tokyo, Kashiwanoha 5-1-5, Kashiwa, Chiba 277-8582 Japan
}

\author{Kenta Hotokezaka}
\affiliation{Research Center for the Early Universe, Graduate School of Science, University of Tokyo, 7-3-1 Hongo, Bunkyo-ku, Tokyo 113-0033, Japan}

\author{Tomoya Kinugawa}
\affiliation{Research Center for the Early Universe, Graduate School of Science, University of Tokyo, 7-3-1 Hongo, Bunkyo-ku, Tokyo 113-0033, Japan}
\affiliation{Faculty of Engineering, Shinshu University, 4-17-1, Wakasato, Nagano-shi, Nagano, 380-8553, Japan}
\affiliation{Research Center for Advanced Air-mobility Systems, Shinshu University, 4-17-1, Wakasato, Nagano-shi, Nagano, 380-8553, Japan}

\date{\today}%

\begin{abstract}
The spins of merging binary black holes (BBHs) inferred from gravitational-wave (GW) observations provide key insights into their formation channels. However, spin parameters are typically weakly constrained from data, and their inferred values are often strongly influenced by the assumed prior in Bayesian analyses. 
A commonly used prior, uniform in spin magnitudes and isotropic in spin directions, assigns vanishing probability density to spin–orbit–aligned configurations, potentially biasing inferences for BBH parameters.
The prior choice can also affect population-level analyses by degrading the convergence of Monte Carlo integrations used to evaluate the likelihood in hierarchical Bayesian inference.
In this work, we propose a novel spin prior that is uniform in the effective spin parameters $\Xeff$ and $\Xp$, two spin combinations that can be relatively well measured from GW data, conditioned on the mass ratio. 
Using simulated BBH populations, we show that the inferred spin population can depend on the choice of prior, and that the proposed prior more accurately recovers the underlying spin population, particularly when the true distribution favors aligned-spin configurations.
Because mass and spin measurements are correlated, our prior also enables a more accurate recovery of the underlying mass distribution.


\end{abstract}

\maketitle

\section{Introduction}
    \label{Intro}
    Gravitational-wave (GW) signals generated by mergers of compact binaries provide information about their masses, spins, and other properties.
    Since the first direct detection of GWs in 2015 \cite{LIGOScientific:2016aoc} by the Laser Interferometer Gravitational-Wave Observatory (LIGO) \cite{LIGO}, the LIGO-Virgo-KAGRA collaboration (LVK) \cite{Virgo, KAGRA, LVK} has observed hundreds of GW signals \cite{GWTC-1, GWTC-2, GWTC-2.1, GWTC-3, LIGOScientific:2025slb}. 
    All the GWs directly detected so far are consistent with signals emitted by mergers of compact binaries, with the majority originating from binary black holes (BBHs). Those observations enabled us to study the statistical properties of merging BBHs, such as the distributions of their masses, spins, and redshifts \cite{GWTC-1_pop, GWTC-2_pop, GWTC-3_pop, GWTC-4_pop}.

    The recovered spin distribution is particularly useful for distinguishing the formation process of BBHs \cite{BBH_formation}. There are two primary scenarios for BBH formation: isolated binary evolution, where two stars evolve together into a BBH system \cite{isolated_formation, isolated_formation_PopIII, isolated_formation_spin, homogeneous}, and dynamical formation, where black holes pair up dynamically in dense stellar environments \cite{dynamically_GC, young_GC, previous_mergers, dynamical_GC_LIGO}. In the isolated binary evolution scenario, the two progenitor stars form from the same molecular cloud core and therefore start with nearly aligned spins. During their binary evolution, binary interactions tend to align the stellar spins with the orbital angular momentum \cite{Tidal_Zahn, Tidal_Hut, Tidal_Kushnir, MT_Packet, homogeneous, PopIIISpin}. If the natal kicks at collapse are modest, this aligned configuration is largely preserved, so the spins of the resulting black holes tend to be preferentially aligned with the orbit \cite{isolated_formation_spin}. In contrast, in the dynamical formation scenario, the black holes originate from unrelated stellar progenitors and acquire their spins independently. The binaries are assembled through multi-body gravitational encounters and exchange interactions, which repeatedly reorient the orbital plane and erase any correlation between the spins and the orbital angular momentum. As a result, the spin directions in dynamically formed systems are expected to be distributed isotropically \cite{BBH_formation, RodriguezGCspin, final_spin}. 

    Among the six components of the two spin vectors, the effective spin parameter is one of the best-measured spin combinations. This parameter is defined as \cite{Xeff_OG}
    \begin{equation}
        \Xeff \coloneqq \frac{m_1\XIz + m_2\XIIz}{m_1+m_2},
    \end{equation}
    where $m_1$ and $m_2$ denote the masses of the heavier and lighter components, respectively, and $\XIz$ and $\XIIz$ are the components of their dimensionless spin vectors projected along the orbital angular momentum.
    This spin combination is conserved during the orbital evolution up to the second Post-Newtonian order \cite{Xeff_conserv}, and its inference does not significantly depend on the reference frequency at which it is computed. 
    
    
    Another spin combination that is relatively well measured and characterizes the spins perpendicular to the orbital angular momentum is the effective precession spin parameter, defined as
    \cite{Xp, Xp_L_eom}:
    \begin{equation}
    \Xp \coloneqq \mathrm{max}\left(\XIp, \frac{3+4q}{4+3q}q\XIIp\right),
    \end{equation}
    where mass ratio $q=m_2/m_1$ and $\XIp$ and $\XIIp$ denote the magnitudes of the spin components lying in the plane perpendicular to the orbital angular momentum.

    The effective spin parameters introduced above serve as useful discriminators of BBH formation pathways. In isolated binary evolution, $\Xeff$ is expected to be preferentially positive and $\Xp$ to be close to zero, reflecting aligned spins with little in-plane spin components. In contrast, for dynamically formed BBHs, the $\Xeff$ distribution is expected to be symmetric about zero, while nonzero values of $\Xp$ are anticipated due to an isotropic spin distribution.

    To infer the population properties of merging BBHs from the GW dataset, we commonly rely on hierarchical Bayesian inference \cite{GWTC-1_pop, GWTC-2_pop, GWTC-3_pop, GWTC-4_pop}. In this scheme, the likelihood consists of the evidence of individual events and a selection function that accounts for the Malmquist bias \cite{Malmquist1922, Malmquist1925}. The evidence of each event is the probability of observing that event under the assumed population, and computing it requires a high-dimensional integration over the BBH source parameters.
    In practice, these integrals are typically evaluated using Monte Carlo integration: first, one performs parameter estimation by Bayesian inference using some fiducial prior distribution, then resultant posterior samples are reweighted to express the integral by sum.
    
    The hierarchical likelihood itself is mathematically independent of the fiducial priors used in the analyses of individual events. However, due to finite sample sizes, the choice of prior may lead to extremely high uncertainty in the evaluation of specific regions within the parameter space. As such, there is a risk that the accuracy of the Monte Carlo integration is severely degraded when a population of BBHs occupies regions that are poorly sampled under the adopted prior.
    In particular, for spin parameters, the commonly adopted prior assumes uniform distributions in spin magnitudes and isotropic distributions in spin directions.
    This prior assigns vanishing measure to configurations in which the spins are exactly collinear with the orbital angular momentum ($\Xp = 0$), and leads to progressively poor constraints in their vicinity. This prior-induced suppression may result in failures of Monte Carlo integration near this region. Consequently, it may bias inferences against the isolated binary evolution scenario, for which the population is expected in this part of parameter space.


    In this work, we propose a novel spin prior that is uniform in $\Xeff$ and $\Xp$, conditioned on $q$. This prior covers the region of parameter space near $\Xp = 0$, while also retaining support in regions where the conventional prior has significant probability density. To assess the impact of the prior choice on recovered population distributions and to demonstrate the utility of the proposed prior, we perform simulation studies in which BBH events are generated from an assumed population model and analyzed using either the conventional spin prior or our new prior. The underlying spin distribution is then recovered using hierarchical Bayesian inference. We find that the choice of spin prior does affect the recovered spin distributions, and that our proposed prior more accurately recovers the true distribution, particularly when it has support near $\Xp = 0$.

    This paper is organized as follows. In Sec.~\ref{method}, we review the basic theory of hierarchical Bayesian inference and describe the spin priors newly adopted in this study. In Sec.~\ref{results}, we perform simulations of population inference and compare the results obtained using this new prior with those obtained using the conventional prior. We first describe the simulation setup in detail. We compare the conventional and novel spin priors for individual-injection parameter estimation, and then examine their differences at the population level. Finally, we conclude in Sec.~\ref{conclusion}. App.~\ref{Calculations_of_prior} provides a detailed description of the newly implemented priors and the calculations of the functions required for sampling from these priors.

\section{formulation}
\label{method}
    \subsection{Hierarchical Bayesian inference}
    \label{method_hierarchical}
    Here we briefly review the hierarchical Bayesian inference in the context of GW astronomy \cite{pop_hyperposterior, pop_inference, GWTC-3_pop, GWTC-4_pop}. This method allows us to estimate information about population properties in a Bayesian manner by incorporating all information about the prior and posterior distributions of parameter estimation for individual events, the assumed population-level distribution of the physical parameters, and the limitations of the detectors. The statistical characteristics of BBH parameter distributions—such as mass, spin, and redshift—provided by this method are crucial for understanding astrophysical BBH formation pathways. Furthermore, as the number of detected signals increases, so does the statistical significance and accuracy of this method.

    In hierarchical Bayesian inference, the goal is to obtain the posterior distribution of the hyperparameters $\bm{\Lambda}$ that describe the population model for BBH parameters $\bm{\theta}$ such as masses, spins, luminosity distance, etc.
    Here, the hyperparameters $\bm{\Lambda}$ can be decomposed into the merger rate $\mathcal{R}$ and the parameters $\bm{\lambda}$ that describe the shape of the population distribution.
    It is known that the merger rate $\mathcal{R}$ can be analytically marginalized by choosing the log-uniform prior distribution for $\mathcal{R}$ \cite{pop_hyperposterior}. Therefore, in cases like this study, where there is no interest in the posterior distribution for $\mathcal{R}$, there is no need to explicitly sample the posterior distribution.
    In such cases, one has to evaluate the posterior distribution of $\bm{\lambda}$ given the observed GW events, denoted as $D$, $p(\bm{\lambda}|D).$ This can be expressed as \cite{pop_hyperposterior, pop_inference, pop_GWTC3}: 

    \begin{align}
        p(\bm{\lambda}|D)=\frac{\pi(\bm{\lambda})}{p(D)}\alpha(\bm{\lambda})^{-N_\mathrm{obs}}
        \prod_{i=1}^{N_\mathrm{obs}} \int \mathrm{d}\bm{\theta}p_\mathrm{pop}(\bm{\theta}|\bm{\lambda})p(d_i|\bm{\theta}),
        \label{hyperposterior}
    \end{align}
    where $d_i$ represents the data of the $i$-th event, and the total number of such events is $N_{\text{obs}}$.
    $\pi(\bm{\lambda})$ represents the hyperprior of the hyperparameters and $p(d_i|\bm{\theta})$ is the likelihood for the $i$-th event. $p(D)$ is the overall normalization factor, also known as the evidence. $\alpha(\bm{\lambda})$ is the selection function, which we will discuss further in the next paragraph. $p_{\text{pop}}(\bm{\theta} | \bm{\lambda})$ is the probability density that describes the population distribution model. 
    For example, if the population model is assumed to be a Gaussian distribution, $p_{\text{pop}}(\bm{\theta} | \bm{\lambda}) = \mathcal{N}(\bm{\theta}; \mu, \sigma^2)$,
    then the hyperparameters $\bm{\lambda}$ are the mean $\mu$ and the variance $\sigma^2$.

    The selection function $\alpha(\bm{\lambda})$ accounts for the detector sensitivity. 
    The selection function is calculated as follows:
    \begin{align}
        \alpha(\bm{\lambda}) \coloneq& \frac{N_\mathrm{det}}{N}\notag\\
        =&\int \mathrm{d}\bm{\theta}p_\mathrm{pop}(\bm{\theta}|\bm{\lambda})p_\mathrm{det}(\bm{\theta}),
        \label{selection_func}
    \end{align}
    where $N$ is the total number of BBHs in the Universe including those that cannot be observed with the current detector sensitivities, $N_{\mathrm{det}}$ is the mean number of detectable signals, and $p_{\mathrm{det}}(\bm{\theta})$ is the detection probability as a function of the parameter $\bm{\theta}$.

    The integral in Eq.~\eqref{hyperposterior} involves integration over $\bm{\theta}$, and in the case of BBH, it typically requires integration over 15 dimensions, making it impossible to perform analytically. Therefore, Monte Carlo integration is usually employed to perform the integration numerically:
     \begin{align}
        \int \mathrm{d}\bm{\theta}p_\mathrm{pop}(\bm{\theta}|\bm{\lambda})p(d_i|\bm{\theta})&\propto
        \int \mathrm{d}\bm{\theta}
        \frac{p_\mathrm{pop}(\bm{\theta}|\bm{\lambda})p(\bm{\theta}|d_i)}{\pi_\mathrm{PE}(\bm{\theta})}\notag\\
        &\approx \frac{1}{N_i}\sum_{j=1}^{N_i}\frac{p_\mathrm{pop}(\bm{\theta}_i^j|\bm{\lambda})}{\pi_\mathrm{PE}(\bm{\theta}_i^j)},
        \label{hyperposterior_numerical}
    \end{align}
    where we use Bayes' theorem to convert $p(d_i|\bm{\theta})$ to $p(\bm{\theta}|d_i).$ This also introduces the prior distribution for parameter estimation, $\pi_\mathrm{PE}(\bm{\theta})$, into the approximation formula.  $\bm{\theta}_i^j, (j=1,\ldots,N_i)$ is the $j$-th sample with respect to the source parameter $\bm{\theta}$, and is assumed to follow the posterior distribution $p(\bm{\theta}|d_i)$ of the $i$-th event. A similar treatment is performed to evaluate the selection function $\alpha(\bm{\lambda})$ as well:
    \begin{align}
        \alpha(\bm{\lambda})&\approx \frac{1}{N_\mathrm{draw}}\sum_{i=1}^{N_\mathrm{found}}\frac{p_\mathrm{pop}(\bm{\theta}_i|\bm{\lambda})}{p_\mathrm{draw}(\bm{\theta}_i)},\label{selection_func_numerical}
    \end{align}
    where $\theta_i$ are samples from some distribution $p_{\text{draw}}(\theta)$. $N_{\text{draw}}$ represents the total number of injections, while $N_{\text{found}}$ denotes the number of detected samples that meet the detection threshold. Note that we only sum up $\theta_i$ that satisfies detection criteria to account for the detector's sensitivity.

    In Monte Carlo integration, an effective sample size $N_\mathrm{eff}$ can be defined in terms of the weights $w_i>0$ in the Monte Carlo summation as follows \cite{SurveySampling}:
    \begin{align}
        N_\mathrm{eff} \coloneq \frac{\left(\sum_{j=1}^{N}w_j\right)^2}{\sum_{j=1}^{N}w_j^2}.
        \label{neff_def}
    \end{align}
    For example, in the case of Eq.~\eqref{selection_func_numerical}, the weight $w_i$ is $p_\mathrm{pop}(\bm{\theta}_i|\bm{\lambda})/p_\mathrm{draw}(\bm{\theta}_i)$.
    $N_\mathrm{eff}$ roughly represents the number of samples that are effectively contributing out of the total number of samples $N$ in Monte Carlo integration, and has a range of $1<N_\mathrm{eff}\leq N$. If the summation does not converge sufficiently, the effective sample size becomes smaller and approaches unity in the limiting case. 
    Therefore, hyperparameter samples that lead to a low effective sample size are considered ineffective and discarded. In practice, the previous studies \cite{pop_GWTC3} and this analysis set the effective sample size threshold to 10 for each event's Monte Carlo integration Eq.~\eqref{hyperposterior_numerical}, and to $4 \times N_\mathrm{event}$ for the selection function Monte Carlo integration Eq.~\eqref{selection_func_numerical}. 
    The rationale behind setting the threshold of $N_\mathrm{eff}$ for this selection effect to this value are discussed in Ref. \cite{n_eff_th}.
    In sufficiently converged Monte Carlo integrations, posterior-sample reweighting provides a reliable approximation to the population likelihood, as quantified by the effective sample size $N_{\rm eff}$.
    In this way, by calculating the posterior with respect to the hyperparameters, population inference can be performed.

    \subsection{Uniform effective spin prior}
    \label{uniform_xexpu_prior_implementation}
\begin{figure}[htb]
        \includegraphics[width=0.8\columnwidth]{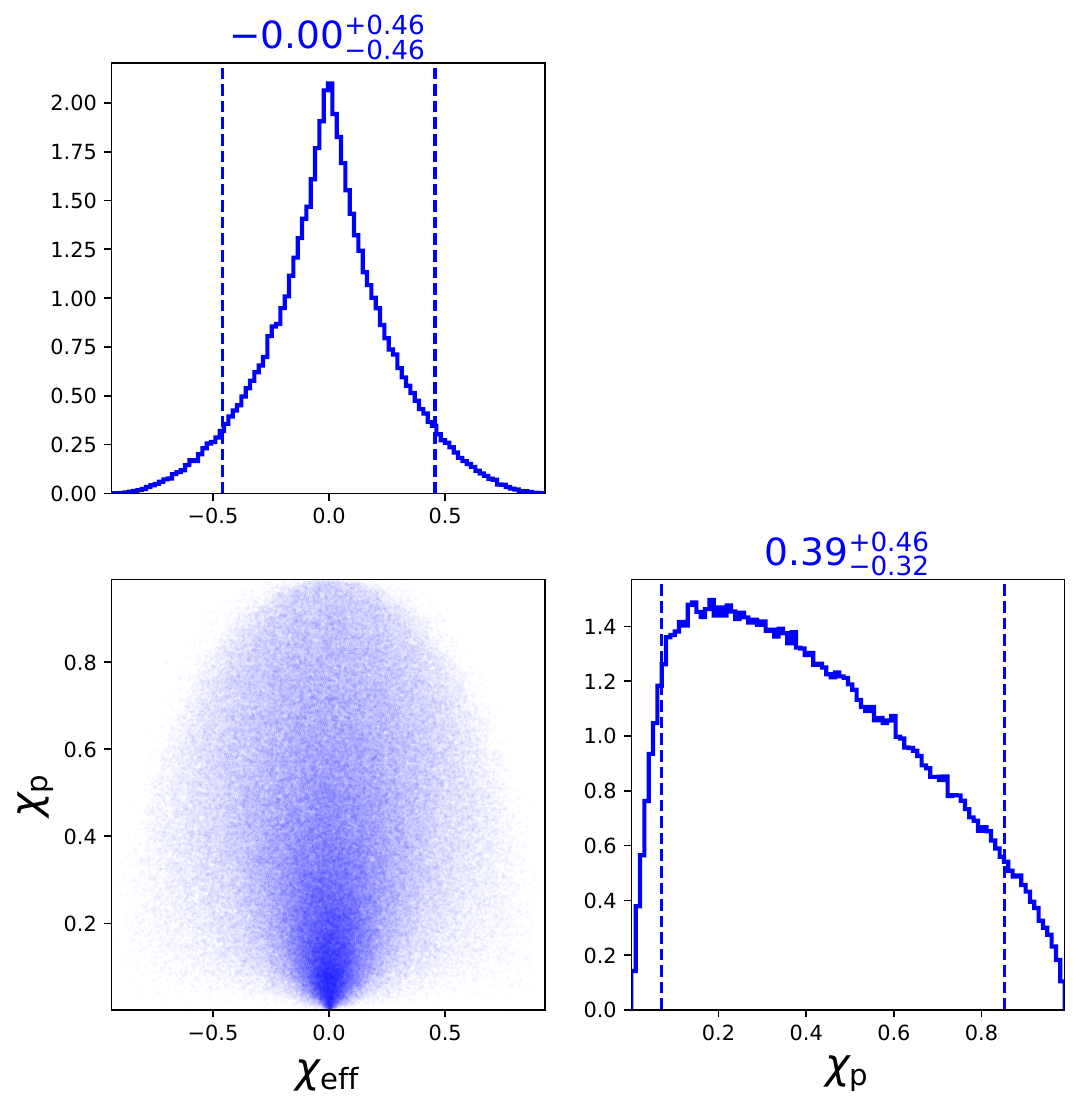}
        \caption{Corner plot of $\Xeff$ and $\Xp$ under the isotropic prior that is used in parameter estimation by the LVK collaboration. The plot is generated from samples drawn according to that prior distribution. 
        Here, the prior on $q$ is chosen so that $m_1$ and $m_2$ are uniformly distributed, which gives a prior on $q$ proportional to $(1+q)^{2/5} / q^{6/5}$ \cite{Thesaurus_prior}. In addition, 
        due to the constraint of dimensionless spin magnitude $a\leq 1$, the allowed region in the $\Xeff$–$\Xp$ plane takes an approximately elliptical shape. The explicit form of the boundary is given by Eq.~\eqref{X_p_max}.
        The dashed line shows 90\% confidence line.
        }
        \label{prior_amp_theta}
    \end{figure}

    Regarding the prior distribution of spin used in the LVK collaboration's analyses \cite{GWTC-2.1, GWTC-3}, the uniform distribution for the spin magnitudes $a_1, a_2$ and the isotropic angle distribution are assumed, in other words, a sine distribution for the angles $\theta_1, \theta_2$ between the spin and the z-axis. Hereafter, we refer to this prior distribution as the isotropic prior distribution. 
    The probability distribution of $\Xeff$ and $\Xp$ under the isotropic prior distribution is shown in Fig.~\ref{prior_amp_theta}, and has a complicated analytic form \cite{analytic_prior}. With this distribution, the prior probability of the effective spins is small near $\Xp \approx 0$.

    Theoretically, the choice of prior distribution in individual Bayesian inference does not pose fundamental issues for hierarchical inference. Practically, when working with finite posterior samples in hierarchical Bayesian inference, this choice can be consequential: for population analyses of effective spin parameters, the isotropic prior can cause divergence in spin regions around $\Xp \approx 0$ because of the tiny denominator in Eq.~\eqref{hyperposterior_numerical}.
    If this happens, the evaluation of the hyperposterior distribution that favors these regions will be discarded only because of the use of a certain prior distribution for parameter estimation.

    Given the issue, in this study, we propose a new spin prior distribution whose marginal distribution for $\Xeff$ and $\Xp$ is uniform. Hereafter, we refer to this prior distribution as the $\Xeff$-$\Xp$ uniform prior distribution.
    This new prior is designed to be uniform within the region where both parameters are physically allowed. Because the extent of this region depends on $q$ \cite{analytic_prior}, this prior probability distribution is expressed as a conditional distribution with respect to $q$. We implemented this prior into \texttt{BILBY} \cite{bilby}, a Bayesian parameter estimation package employed by the LVK collaboration.
    
\begin{figure}[htb]
        \includegraphics[width=0.8\columnwidth]{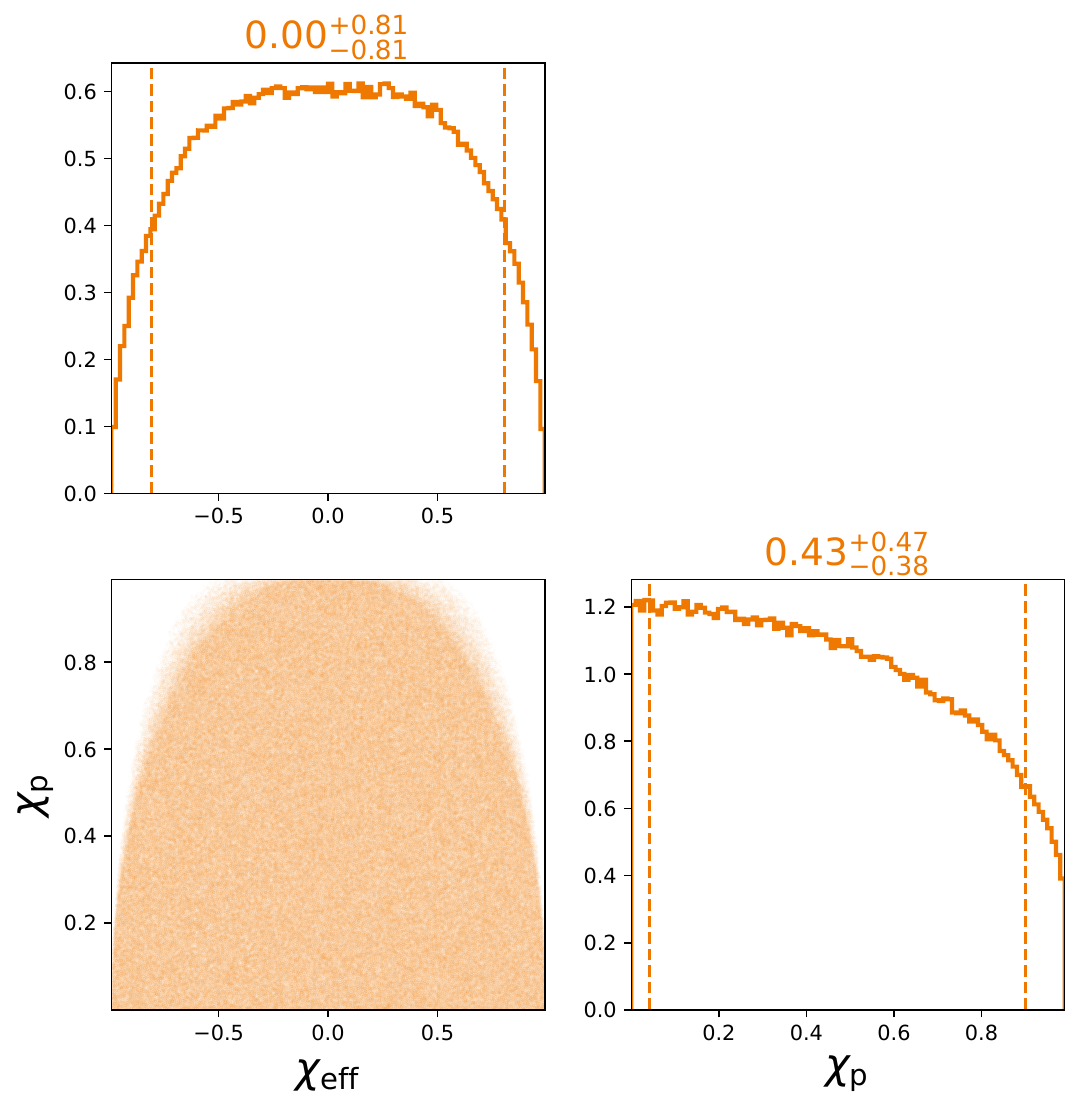}
        \caption{Corner plot of $\Xeff$ and $\Xp$ under the $\Xeff$-$\Xp$ uniform prior. The plot is generated from samples drawn according to the prior distribution of Eq.~\eqref{Xe_Xp_uni_prior}.
        The plotting procedure is the same as in Fig.~\ref{prior_amp_theta}. Although this prior is uniform over the allowed region in the $\Xeff$–$\Xp$ plane, it does not appear uniform near the boundary in the $\Xeff$–$\Xp$ plane because the allowed region itself depends on $q$, which is not fixed here.
        }
        \label{xexpuni_prior}
    \end{figure}

    Fig.~\ref{xexpuni_prior} shows the $\Xeff$-$\Xp$ distribution of $\Xeff$-$\Xp$ uniform prior distribution. Note that this is not a simple box-shaped uniform distribution. This is because the absolute value of each spin parameter cannot exceed unity, thereby constraining the values that $\Xeff$ and $\Xp$ can simultaneously take. The functional form of this prior is given as follows:
    \begin{align}
        &\pi_\mathrm{PE}\left(\Xeff, \Xp|q\right)\notag\\
        &= N_\chi(q)\Theta\left(|\Xeff|< 1\right)
        \Theta\left(0<\Xp< \chi_\mathrm{p\_max}\left(q,\Xeff\right)\right),
        \label{Xe_Xp_uni_prior}
    \end{align}
    where $\Theta$ is the Heaviside step function and $\chi_\mathrm{p\_max}(q, \Xeff)$ denotes the physically maximum value of $\Xp$ for given $\Xeff$ and $q$ \cite{analytic_prior}:
    \begin{align}
        &\chi_\mathrm{p\_max}(q, \Xeff) \notag \\
            &=
            \begin{cases}
            1, & |\Xeff| \leq \dfrac{q}{1+q}, \\
            \sqrt{1-\left((1+q)|\Xeff|-q\right)^2}, & |\Xeff| > \dfrac{q}{1+q},
            \end{cases}
        \label{X_p_max}
    \end{align}
    and $N_\chi(q) = (2+2q)/(\pi+4q)$ is the normalization factor of this distribution. 
    This prior can alternatively be expressed as:
    \begin{align}
        \pi_\mathrm{PE}(\Xeff,\Xp|q)&=N_\chi(q)\Theta(0<\Xp<1)\notag\\
        &\quad\times\Theta\qty(\abs{\Xeff}<\frac{q+\sqrt{1-\Xp^2}}{1+q}).
        \label{Xe_Xp_uni_prior_alt}
    \end{align}
    The marginal distribution for $\Xeff$, conditioned by $q$, is given by
    \begin{align}
        \pi_\mathrm{PE}(\Xeff|q)=N_\chi(q)\chi_\mathrm{p\_max}(q, \Xeff).
    \end{align}
    On the other hand, the marginal distribution for $\Xp$, conditioned by $q$, is expressed in a much simpler form:
    \begin{align}
        &\pi_\mathrm{PE}(\Xp|q) = \frac{4}{\pi+4q}\qty(q+\sqrt{1-\Xp^2}).
    \end{align}
    The procedures to derive these formulae are provided in App.~\ref{Calculations_of_prior}.

    Note that, however, there are infinite ways to select a spin prior distribution such that the marginal distribution for the  $\Xeff$-$\Xp$ plane is uniform. This is because the condition constrains only two degrees of freedom out of six. 
    In this study, we implement a conditional prior in which, for given $\Xeff$, $\Xp$, and $q$, $\XIz$ and $\XIIz$ are uniformly distributed over their allowed parameter region. Furthermore, for given $\Xeff$, $\Xp$, $q$, $\XIz$, and $\XIIz$, the conditional prior for $\XIp$ and $\XIIp$ is taken to be proportional to $\XIp\times\XIIp$ within their allowed region.
    The sampling method for all spin parameters from this distribution is described in App.~\ref{Calculations_of_prior}.

\begin{figure}[htb]
        \includegraphics[width=0.8\columnwidth]{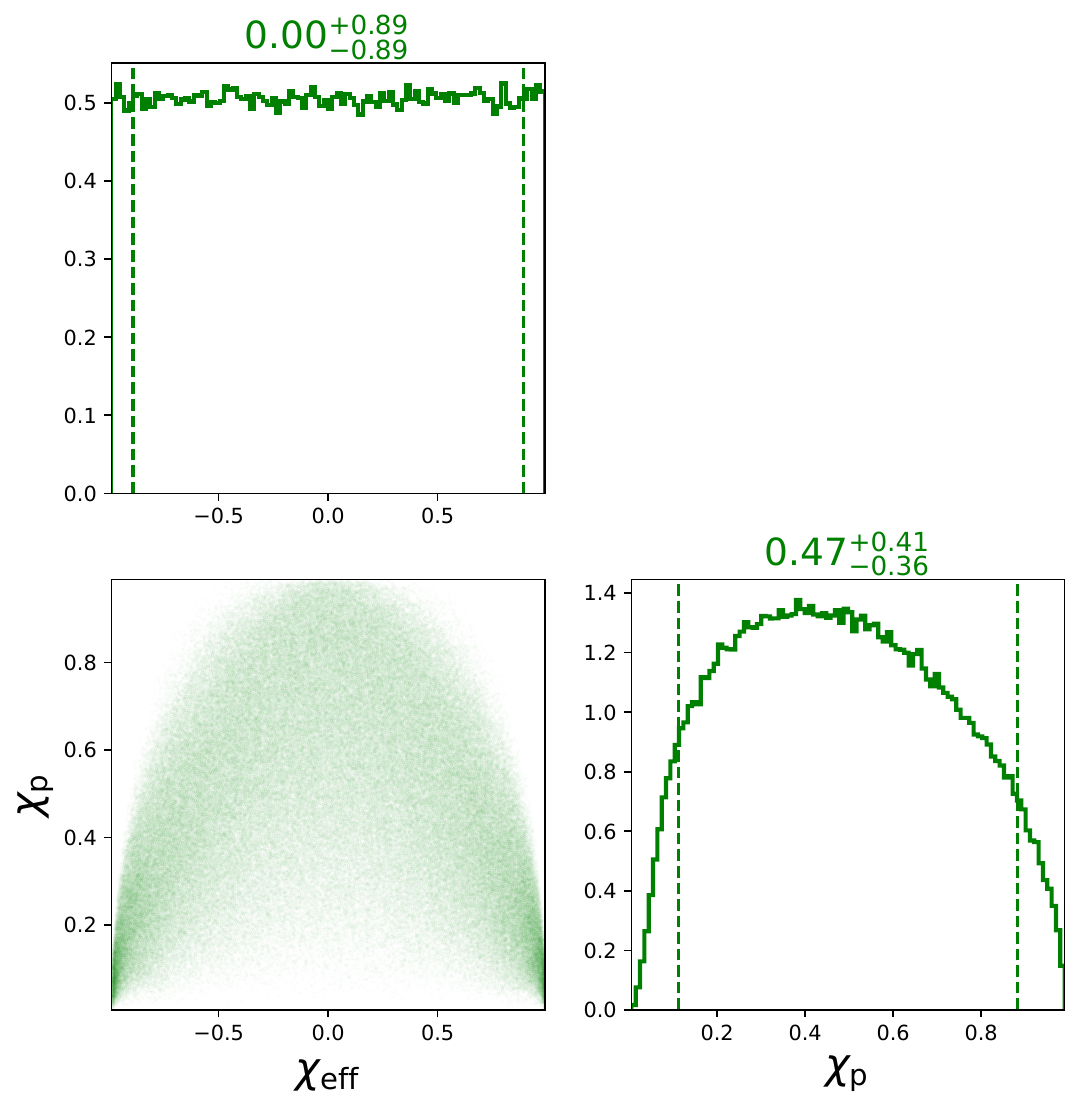}
        \caption{Corner plot of $\Xeff$ and $\Xp$ under the only $\Xeff$ uniform prior. The plot is generated from samples drawn according to the prior distribution (see Eq. (2) of Ref.~\cite{Xeff_uniform}). 
        The plotting procedure is the same as in Fig.~\ref{prior_amp_theta}.}
        \label{only_xeuni_prior}
    \end{figure}

     We briefly mention a prior distribution proposed in a previous work by Zackay et al. \cite{Xeff_uniform}.
     This prior distribution is designed such that the marginal distribution of $\Xeff$ becomes uniform, as shown in Fig.~\ref{only_xeuni_prior}. This prior has a broadly distributed prior probability density for $\Xeff$, making it suitable for analyses that focus solely on $\Xeff$. However, 
     the probability density under this prior is nearly zero when both $\Xeff$ and $\Xp$ are close to zero. We find that this behavior leads to biased population inference when the true spin distribution is concentrated in this region. We therefore adopt our $\Xeff$–$\Xp$ uniform prior instead. 

\section{simulation}
\label{results}

    \subsection{Settings}
    \label{settings}  
    We perform simulations with mock populations to assess the impact on both individual parameter estimation and hierarchical analysis caused by changing the prior. The procedure involves assuming simulated populations, followed by sampling BBH injection parameters based on the simulated populations. The injection data is analyzed using both the isotropic prior distribution and the uniform $\Xeff$-$\Xp$ prior. Subsequently, we perform hierarchical Bayesian inference using the resulting posterior distribution samples.
    
    The functional form of simulated populations is identical to that used in previous studies \cite{pop_GWTC3}: a distribution model consisting of a power-law plus a peak component in mass and a two-dimensional Gaussian distribution in effective spins.
    Specifically, the population-level distribution can be described as:
    \begin{align}
        p_\mathrm{pop}(\bm{\theta}|\bm{\lambda}) &= p_\mathrm{pop}(m_1|\bm{\lambda_{m_1}})p_\mathrm{pop}(m_2|m_1, \bm{\lambda_{m_2}})\notag\\
        &\quad\times p_\mathrm{pop}(z|\lambda_{z})p_\mathrm{pop}(\Xeff, \Xp|\bm{\lambda_{\mathrm{spin}}}),
    \end{align}
    where each part is defined as follows. 
    The part related to $m_1$ is:
    \begin{align}
        &p_\mathrm{pop}(m_1|\bm{\lambda_{m_1}}) \notag\\
        &= p_\mathrm{pop}(m_1|\alpha, M_\mathrm{max}, M_\mathrm{min}=5M_{\odot}, \mu_m, \sigma_m, f_\mathrm{peak})\notag\\
        &=(1- f_\mathrm{peak})\frac{(1+\alpha)m_1^\alpha}{M_\mathrm{max}^{1+\alpha} - M_\mathrm{min}^{1+\alpha}}\notag\\
        &\quad+ f_\mathrm{peak}\frac{1}{\sqrt{2\pi \sigma_m^2}}\mathrm{exp}\left(-\frac{(m_1-\mu_m)^2}{2\sigma_m^2}\right),
        \label{p_pop_m1_func}
    \end{align}
    where $M_\mathrm{min} < m_1 < M_\mathrm{max}$ and it is zero outside this range.
    For $m_2$, it is:
    \begin{align}
        p_\mathrm{pop}(m_2|m_1, \bm{\lambda_{m_2}})  &= p_\mathrm{pop}(m_2|m_1, \beta_q, M_\mathrm{min}=5M_{\odot})\notag\\
        &= \frac{(1+\beta_q)m_2^{\beta_q}}{m_1^{1+\beta_q} - M_\mathrm{min}^{1+\beta_q}},
    \end{align}
    and is zero outside the range $M_\mathrm{min} < m_2 < \min(M_\mathrm{max}, m_1)$.
    Converting $m_2$ into $q = m_2 / m_1$, we see that $\beta_q$ is also the power of $q$.

    The distribution for redshift $z$ is assumed to be:
    \begin{align}
        p_\mathrm{pop}(z|\lambda_{z}) &= p_\mathrm{pop}(z|\kappa)\notag\\
        &\propto \frac{\mathrm{d}V_c}{\mathrm{d}z}(1 +z)^{\kappa -1},
    \end{align}
    where ${\mathrm{d}V_c}/{\mathrm{d}z}$ is the differential comoving volume with respect to redshift $z$, calculated under the Planck15 cosmology \cite{Planck15}.
    This distribution is based on the assumption that the merger rate per unit comoving volume and per unit source-frame time, accounting for time dilation, is proportional to $(1+z)^\kappa$.

    For the spins, we employ the following Gaussian model for effective spin parameters \cite{Miller:2020zox}:
    \begin{align}
        &p_\mathrm{pop}(\Xeff, \Xp|\bm{\lambda_{\mathrm{spin}}}) \notag\\
        &= p_\mathrm{pop}(\Xeff, \Xp|\mu_\mathrm{eff}, \sigma_\mathrm{eff}, \mu_\mathrm{p}, \sigma_\mathrm{p}, \rho)\notag\\
        &\propto \mathcal{N}_\mathrm{2D}(\Xeff, \Xp;\mu_\mathrm{eff}, \sigma_\mathrm{eff}, \mu_\mathrm{p}, \sigma_\mathrm{p}, \rho)\notag\\
        &\quad\times\Theta(-1<\Xeff<1)\Theta(0<\Xp<1),
    \end{align}
    where $\mathcal{N}_\mathrm{2D}$ represents a two-dimensional Gaussian distribution. The distribution for effective spin parameters is normalized so that the distribution has a non-zero value only on $-1<\Xeff<1$ and $0<\Xp<1$.
    Finally, we have 12 hyperparameters: 
    \begin{equation}
        \bm{\lambda}=(\alpha, M_\mathrm{max}, \mu_m, \sigma_m, f_\mathrm{peak}, \beta_q, \kappa, \mu_\mathrm{eff}, \sigma_\mathrm{eff}, \mu_\mathrm{p}, \sigma_\mathrm{p}, \rho).
    \end{equation}
    
    In this study, we generate two simulated astrophysical BBH population distributions. One of them is a GWTC-3–like population whose effective spin distribution is concentrated around $\Xeff \approx 0$ and $\Xp \approx 0.1$, while the other simulated BBH population has an effective spin distribution concentrated in the region of large $\Xeff$. We call them the GWTC-3–like population and the large-$\Xeff$ population, respectively. The values of hyperparameters for these populations are summarized in the Table~\ref{population_hyperparameters}. In both cases, the values of the hyperparameters related to the mass and redshift are identical to the maximum-likelihood values inferred from the GWTC-3 population analysis \cite{pop_GWTC3}. For the GWTC-3–like population, the means and variances of $\Xeff$ and $\Xp$ are chosen to approximate their inferred distributions (Fig. 16 of Ref.~\cite{pop_GWTC3}). 
    By contrast, the large-$\Xeff$ population is inconsistent with the GWTC-3 population results. 
    Although this population also serves as a stress test, it is motivated by the possibility of astrophysical aligned/high-spin subpopulations \cite{large_xeff_subpop, hierchical_xeffxp_subpop}.
    For simplicity, the correlation between $\Xeff$ and $\Xp$, $\rho$, is set to zero for both cases.
    
    Because the GWTC-3–like spin distribution is concentrated in a region of parameter space that is well covered by the isotropic spin prior, the bias that may arise in population inference due to the use of the isotropic prior is expected to be limited.
    On the other hand, the spin distribution of the large-$\Xeff$ population is concentrated around the region where $\Xeff$ is sufficiently large while $\Xp$ is near-zero. Consequently, the use of the isotropic prior is expected to introduce biases in the population inference.
    
    \begin{table}[htb]
        \caption{Hyperparameters of assumed BBH populations. The values from $\alpha$ to $\kappa$ are identical for both cases.}
        \label{population_hyperparameters}
        \vspace{5pt}
        \setlength{\tabcolsep}{6pt}
        \renewcommand{\arraystretch}{1.2}
        \centering
        \begin{tabular}{ccc}
            \hline \hline
            Hyperparameters &  GWTC-3–like & Large $\Xeff$\\ \hline
            $\alpha$ & -3.09 & \\
            $M_\mathrm{max}$ & 92.5 &\\
            $\mu_m$ & 34.1 &\\
            $\sigma_m$ & 1.01 &Same as in the left\\
            $f_\mathrm{peak}$ & 0.0591 &\\
            $\beta_q$ & 1.99 &\\
            $\kappa$ & 1.95 &\\
            $\mu_\mathrm{eff}$ & 0.0602 & 0.700\\
            $\sigma_\mathrm{eff}$ & 0.112 & 0.100\\
            $\mu_\mathrm{p}$ & 0.139 & 0.00\\
            $\sigma_\mathrm{p}$ & 0.193 & 0.100\\
            $\rho$ & 0.00 & 0.00\\
            \hline
        \end{tabular}
    \end{table}
    
    We draw BBH parameters from these populations and perform parameter estimation only for those with an optimal signal-to-noise ratio (SNR) exceeding 10. The number of injections is 70.
    We assume detections by three interferometers: LIGO Hanford, LIGO Livingston, and Virgo, and their noises are assumed to be stationary and Gaussian. The random seed used for noise generation is varied between different injections, but is fixed across analyses with different priors for a given injection, ensuring that the input data are identical. We use a representative O3 power spectral density (PSD) \cite{GWTC-2, O3_released_psd_data}.
    The approximant waveform model used for both the injections and the analyses is consistently set to \texttt{IMRPhenomXPHM}~\cite{IMRPhenomXPHM}.

    The settings for the parameter estimation are as follows.
    The parameter estimation is performed in the full 15-dimensional parameter space. The priors for spins are the isotropic prior and the $\Xeff$-$\Xp$ uniform prior described in Sec.~\ref{uniform_xexpu_prior_implementation}. The maximum value of the spin magnitude is set to 0.99. For simplicity, the prior on the luminosity distance is taken to follow a power law with an index of 2, while the priors for the other parameters are set to the default values used in GWTC-3 analysis \cite{GWTC-3}.
    We use the \texttt{Python} package \texttt{BILBY} \cite{bilby} to perform parameter estimation via nested sampling \cite{nested_john_skilling, dynesty}.

    When performing hierarchical Bayesian inference, it is necessary to evaluate the parameter estimation level prior $\pi_{\mathrm{PE}}$ for the mass and spin parameters, as shown in Eq.~\eqref{hyperposterior_numerical}. For the calculation of the effective spins prior under the isotropic prior, we use the analytic formula derived in Ref.~\cite{analytic_prior}.

    To estimate the selection function for the GWTC-3–like population, we adopt the same $p_{\mathrm{draw}}$ used in the GWTC-3 analysis (see Table XII of Ref. \cite{GWTC-3}), which assumes isotropic spin orientations. We extend the upper redshift bound from 1.9 to 3 to include higher-redshift injections.
    On the other hand, in the analysis of the large-$\Xeff$ population, there are not enough samples in the high $\Xeff$ region to properly evaluate the selection effects.
    Therefore, we modify $p_\mathrm{draw}$ only for the spin parameters to follow the $\Xeff$-$\Xp$ uniform distribution. 
    For a given population model, the same $p_{\mathrm{draw}}$ is used for selection-function estimation in both analyses; the only difference is the prior assigned to each individual event.
    For the summation in Eq.~\eqref{selection_func_numerical}, The number of injections is determined to achieve $N_\mathrm{found}=2\times10^{5}$.
    
    As discussed in Sec.~\ref{method_hierarchical}, the threshold for the effective sample size $N_\mathrm{eff}$ is set to be the same as in the GWTC-3 analysis \cite{pop_GWTC3}. 
    These calculations are carried out using Markov chain Monte Carlo sampling with the \texttt{Python} package \texttt{emcee} \cite{emcee}, based on the code employed in the GWTC-3 population analysis \cite{pop_GWTC3, GWTC3_pop_code}.

    \subsection{Individual injections}
    \label{result_individual_injections}
\begin{figure*}[!htbp]
        \includegraphics[width=1.9\columnwidth]{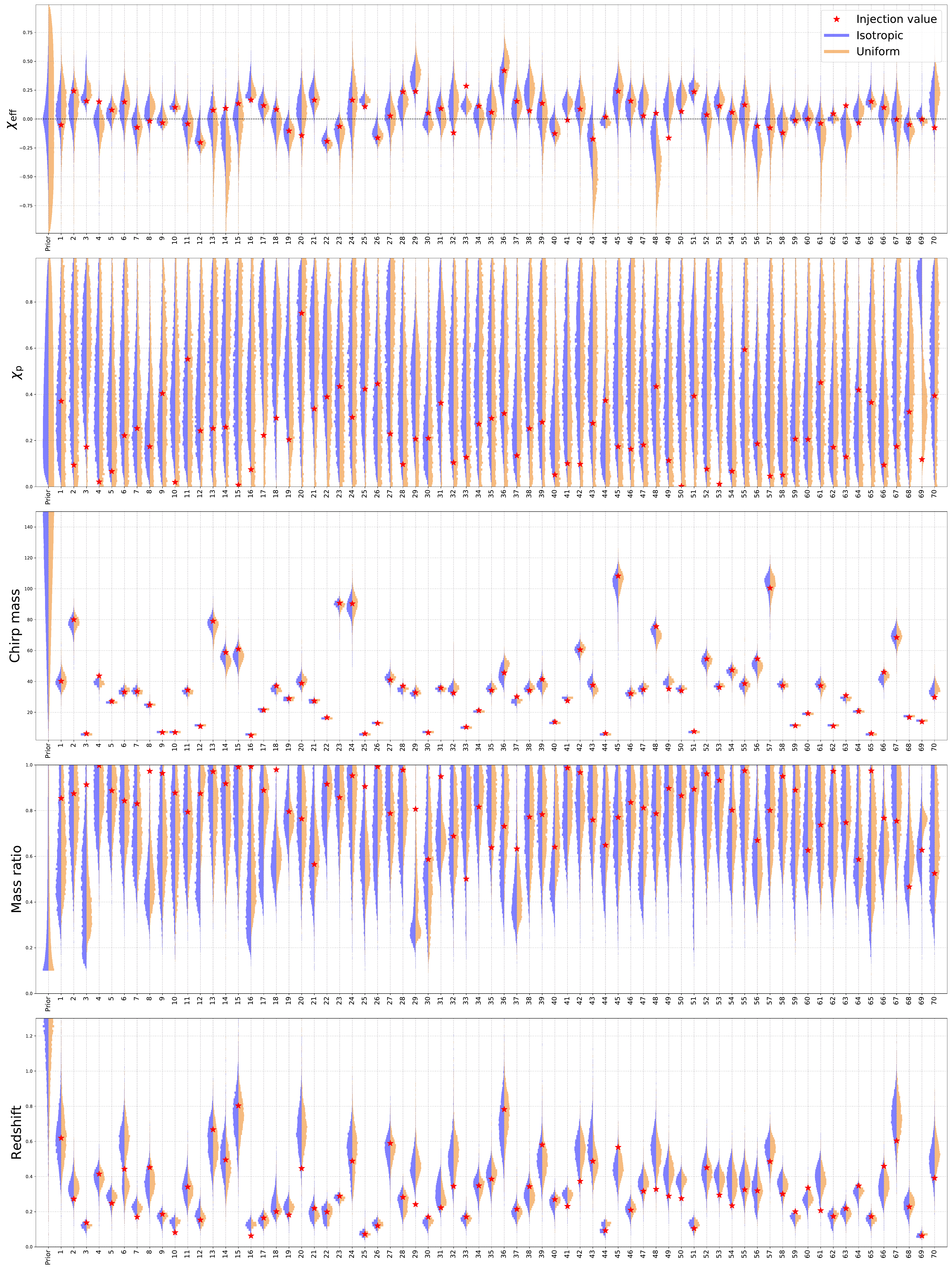}
        \caption{Prior and posterior distributions of the effective spins, chirp mass, mass ratio and redshift. The leftmost distribution corresponds to the prior distribution, while each of the distributions to the right corresponds to the results obtained from analyses of the corresponding injections, where the injection parameters were drawn from the GWTC-3–like population. The red stars represent injection values.
        The blue distribution on the left represents the parameter estimation results using an isotropic prior, which is used in the LVK analysis.
        The orange distribution on the right represents the results using $\Xeff$-$\Xp$ uniform prior.}
        \label{XeXp_GWTC3like_violin}
    \end{figure*}

\begin{figure*}[!htbp]
        \includegraphics[width=1.9\columnwidth]{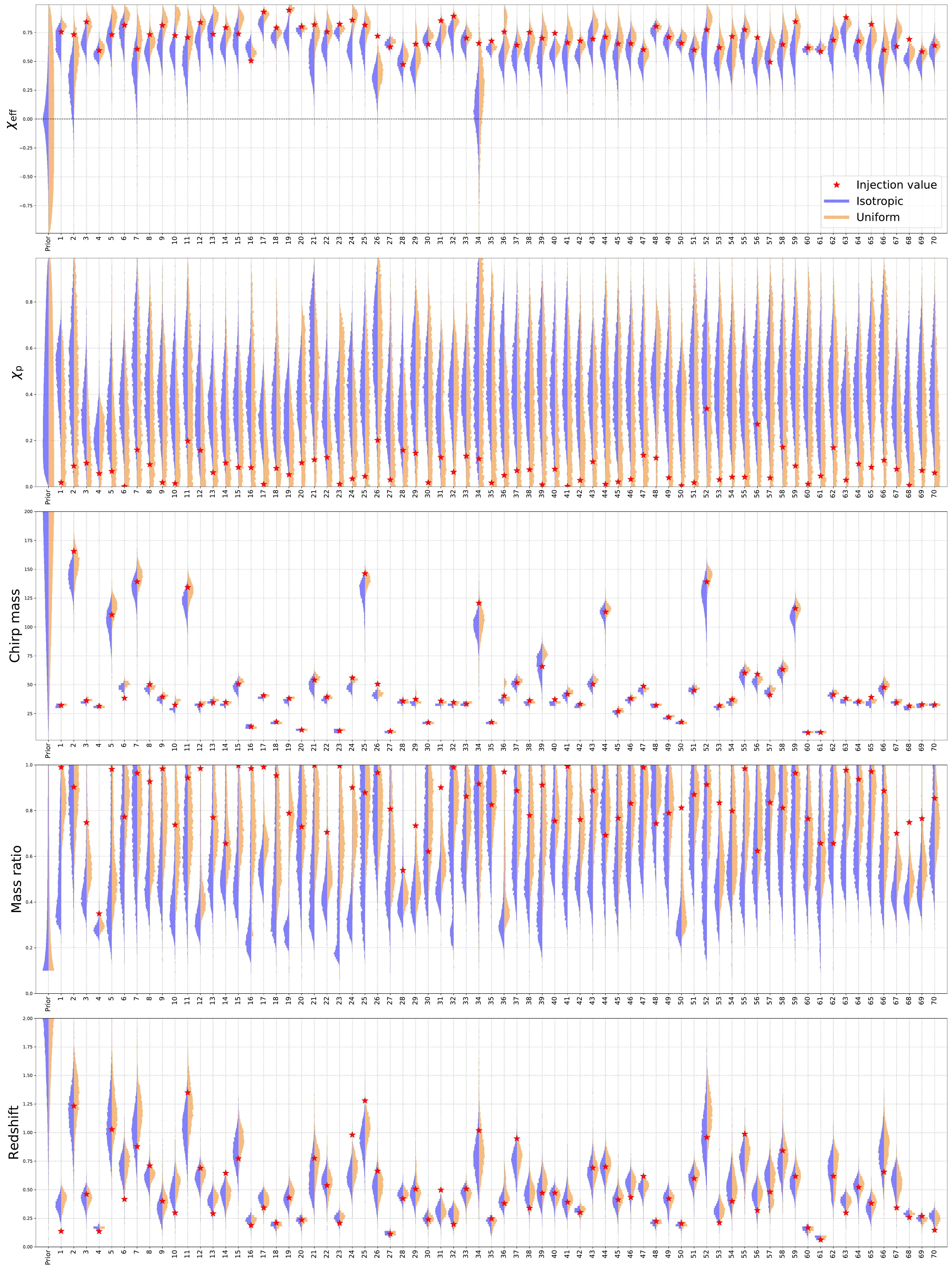}
        \caption{Prior and posterior distributions of the effective spins, chirp mass, mass ratio and redshift for each injection from large-$\Xeff$ population. The figure settings and color scheme follow those used in Fig.~\ref{XeXp_GWTC3like_violin}.}
        \label{XeXp_largexeff_violin}
    \end{figure*}

\begin{figure*}[!htbp]
        \includegraphics[width=1\columnwidth]{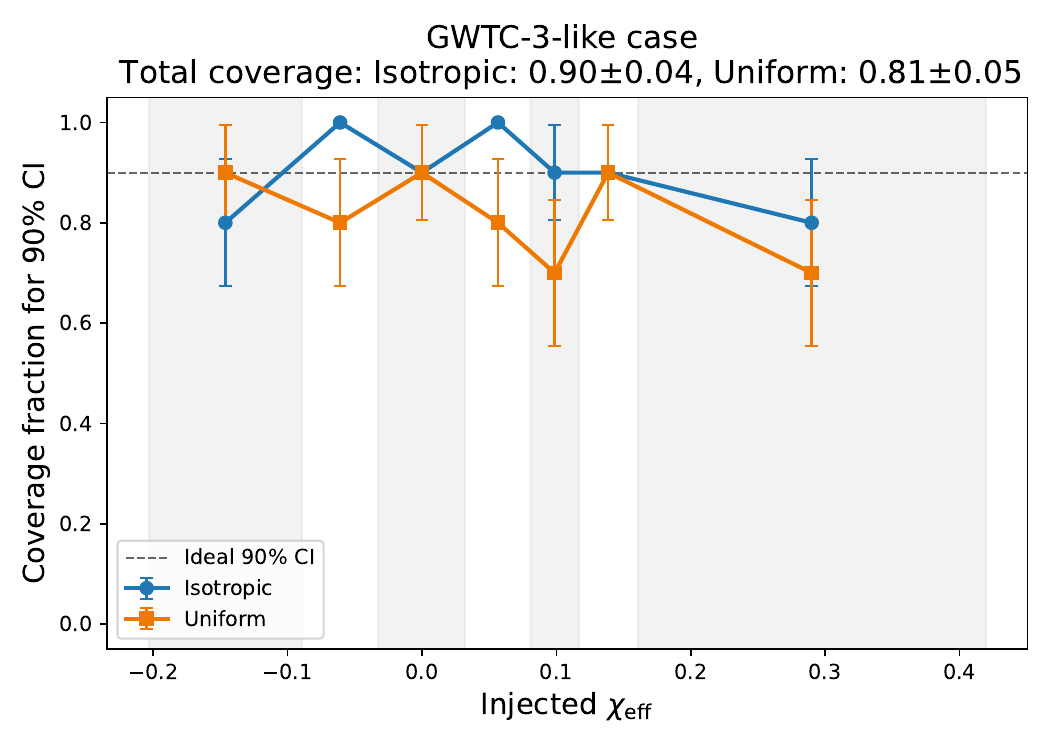}
        \includegraphics[width=1\columnwidth]{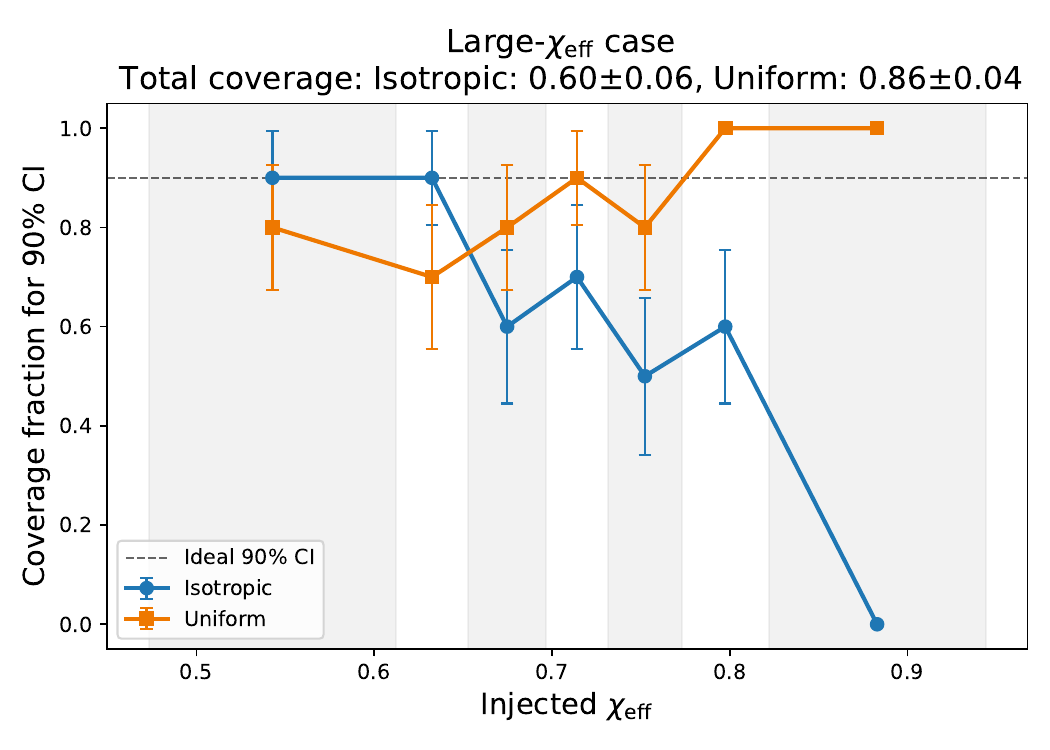}
        \caption{
        Coverage fraction of the injected value as a function of the injected $\Xeff$. Injections are grouped into equal-count bins sorted by the injected $\Xeff$. For each bin, we show the fraction of injections whose true value lies within the 90\% posterior credible interval (CI). The left and right panels show the GWTC-3-like and large-$\Xeff$ populations, respectively. Error bars indicate binomial standard errors. The dashed horizontal line marks the ideal 90\% coverage. Blue and orange curves correspond to the isotropic spin and $\Xeff$-$\Xp$ uniform priors, respectively.}
        \label{coverage_plot_xeff}
    \end{figure*}

    We present the results of parameter estimation for individual injections using the isotropic and $\Xeff$-$\Xp$ uniform priors.
    Fig.~\ref{XeXp_GWTC3like_violin} and Fig.~\ref{XeXp_largexeff_violin} shows the posterior distributions of the effective spins, chirp mass, mass ratio and redshift for 70 injections from the GWTC-3-like and large-$\Xeff$ populations, for both priors. 
    Fig.~\ref{coverage_plot_xeff} compares the coverage fraction of the injected value as a function of the injected $\Xeff$ for the GWTC-3-like and large-$\Xeff$ populations.

    For the GWTC-3-like population, almost all of the posteriors recover their injections for the isotropic prior case and the $\Xeff$-$\Xp$ uniform prior case. 
    For example, when comparing the cases for the isotropic prior and the $\Xeff$-$\Xp$ uniform prior, the fractions of injections whose true values fall within the 90\% credible intervals of the posteriors are 90$\pm$ 4\% and 81$\pm 5$\% for $\Xeff$. The left panel of Fig.~\ref{coverage_plot_xeff} shows the coverage fraction as a function of $\Xeff$. For $\Xp$, the corresponding fractions are 70$\pm$5\% and 90$\pm$4\%, respectively.
    The lower coverage for $\Xp$ under the isotropic prior arises because injections with $\Xp$ values close to zero are disfavored by the isotropic prior.
    Overall, the $\Xeff$ posterior distributions are broader for the $\Xeff$-$\Xp$ uniform prior. This is because the $\Xeff$-$\Xp$ uniform prior assigns higher probability over a wider range in $\Xeff$ compared with the isotropic prior.
    Because masses and spins are correlated in the posterior \cite{mass_spin_correlation, principal_component_analysis}, the measured masses weakly depend on the prior choice; for example, this effect is visible in the mass-ratio posterior for injection 40.
    In contrast, the redshift posteriors are largely insensitive to the choice of prior.

    In most of the injections, the injected values lie within the 90\% credible intervals of their posterior distributions. The true values of $\Xeff$ fall outside the 90\% credible intervals in 7 injections only when using the $\Xeff$–$\Xp$ uniform prior, in 1 injection only when using the isotropic prior, and in 6 injections for both prior choices. In these cases, the true values remain well within the overall posterior support. For this reason, we conclude that those cases are due to noise fluctuations and both priors perform well for this population.

    On the other hand, for the large-$\Xeff$ population case shown in Fig.~\ref{XeXp_largexeff_violin}, the posteriors obtained using the uniform $\Xeff$-$\Xp$ prior more frequently recover the injected values compared to those obtained with the isotropic prior. In fact, when comparing the cases for the isotropic prior and the $\Xeff$–$\Xp$ uniform prior, the fractions of injections whose true values fall within the 90\% credible intervals of the posteriors are $60\pm 6$\% and 86$\pm 4$\% for $\Xeff$. The right panel of Fig.~\ref{coverage_plot_xeff} shows the coverage fraction as a function of $\Xeff$. As can be seen, when the isotropic prior is used, the coverage fraction decreases as the injected $\Xeff$ increases. For $\Xp$, they are 9$\pm 3$\% and 69$\pm 6$\%, respectively.
    In the case of the parameter estimation using an isotropic prior, we find a systematic bias in the posterior distributions. Specifically, the $\Xeff$ posteriors tend to underestimate the injected values, while the $\Xp$ posteriors tend to overestimate them.
    This is because, as shown in Fig.~\ref{prior_amp_theta}, in the region where large-$\Xeff$ population are concentrated (e.g., $\Xeff \gtrsim 0.5$ and $\Xp \lesssim 0.2$), the isotropic prior probability decreases as $\Xeff$ increases and $\Xp$ decreases, and increases in the opposite direction. Therefore, even if the likelihood is large in the region with large $\Xeff$ and small $\Xp$, the constraint imposed by the isotropic prior causes the posteriors to concentrate in the region with smaller $\Xeff$ and larger $\Xp$.

\begin{figure*}[!htbp]
        \includegraphics[width=2\columnwidth]{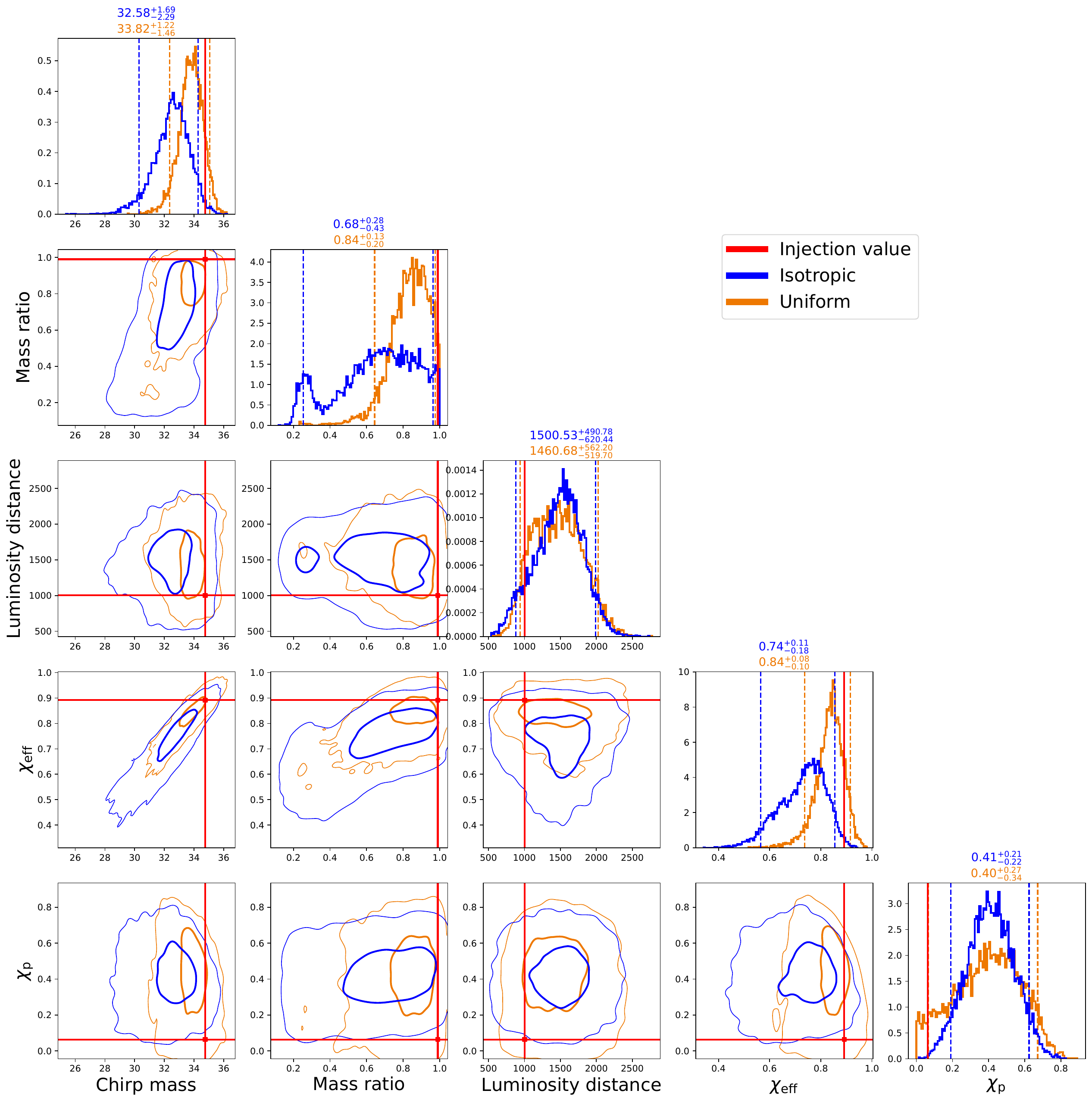}
        \caption{Corner plot of the posterior for the 32nd injection in the large-$\Xeff$ population case. The blue contours represent the posterior obtained using the isotropic prior, the orange contours correspond to the posterior obtained using the $\Xeff$–$\Xp$ uniform prior, and the red lines indicate the injection values. In the plot of the correlations in the off-diagonal components, the thick lines show the contours enclosing 50\% of the total samples, while the thin lines show the contours enclosing 99\% of the total samples.}
        \label{corner_injection_largexeff}
    \end{figure*}

    Since masses and spins are correlated, we also find indications of systematic biases in the posterior distributions of the chirp mass and mass ratio. In particular, 
    the estimated chirp mass and mass ratio under the isotropic prior show a tendency toward underestimation relative to the injected values.
    Fig.~\ref{corner_injection_largexeff} shows a corner plot of the posterior for a typical injection (the 32nd injection) from the large-$\Xeff$ population. As can be seen from this figure, $\Xeff$ is positively correlated with the chirp mass and mass ratio. Given this positive correlation, the underestimation of $\Xeff$ leads to underestimation of chirp mass and mass ratio. The positive correlation between $\Xeff$ and mass ratio observed here contrasts with earlier studies based on gravitational waveforms computed using the post-Newtonian expansion, which reported a negative correlation between aligned spin components and mass ratio. By contrast, our result is consistent with simulation studies assuming black holes with nearly extremal spins \cite{extremal_BH} (see Fig.~3 of that reference).
    
    

    \subsection{Population inference}
\begin{figure*}[htb]
        \includegraphics[width=2\columnwidth]{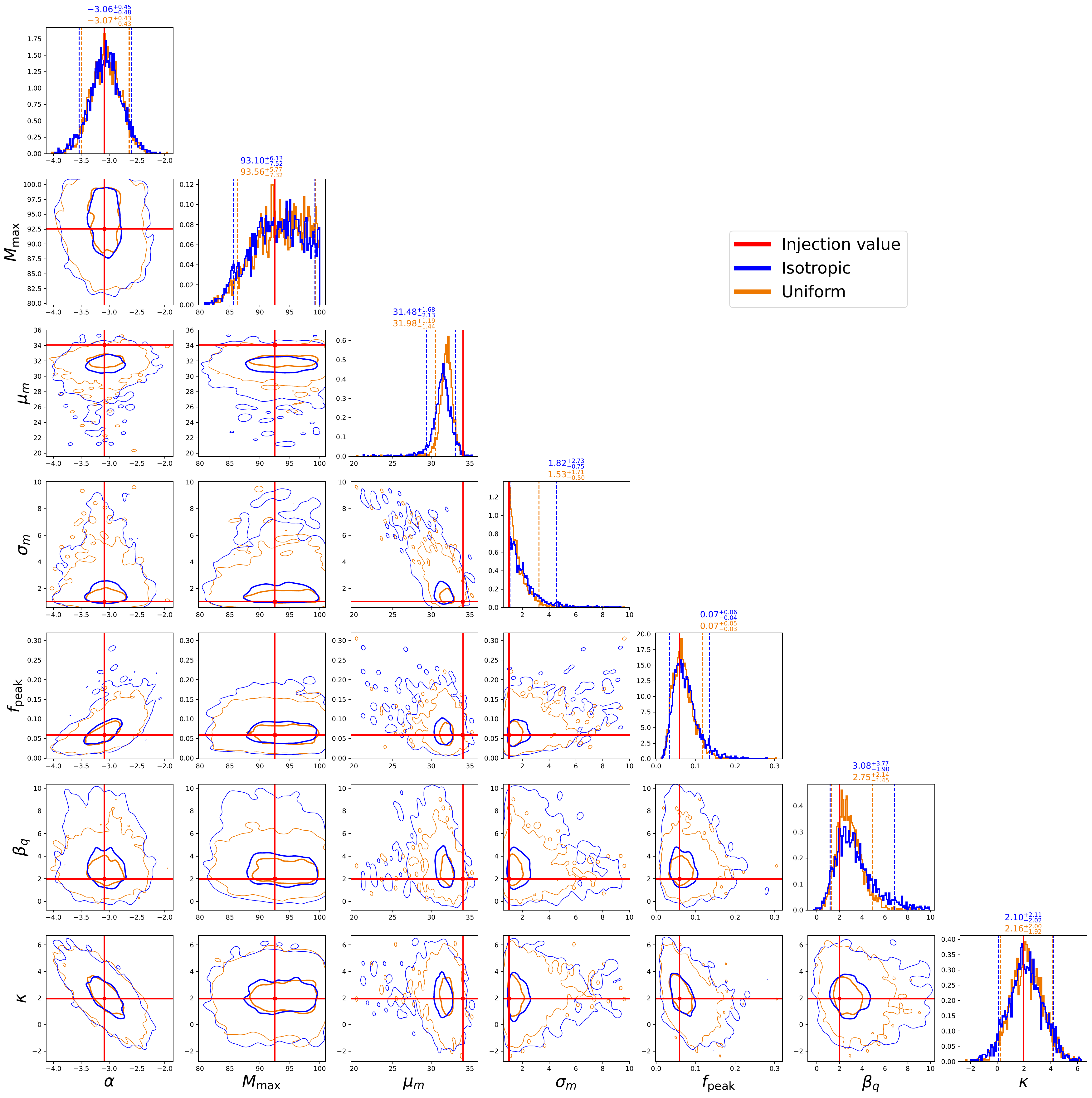}
        \caption{Corner plot for posterior of hyperparameter related to masses and redshift for GWTC-3-like population case. The population distribution model consisting of a power-law plus a peak component in mass and a power-law in redshift. Specifically, $\alpha$ is the power-law index of $m_1$, $M_\mathrm{max}$ is maximum mass, $\mu_\mathrm{m}$ and $\sigma_\mathrm{m}$ are the mean and standard deviation of the Gaussian peak of $m_1$, $f_\mathrm{peak}$ is the fraction of the Gaussian peak in the power-law–plus–peak model for $m_1$, $\beta_q$ is the power-law index of $m_2$ (or $q$), and $\kappa$ is the power-law index of redshift. For details, see Sec.~\ref{settings}. The blue and orange curves correspond to the results obtained with the isotropic prior and the $\Xeff$-$\Xp$ uniform prior, respectively, while the red line indicates the hyperparameter values of the assumed population. In the plot of the correlations in the off-diagonal components, the thick lines show the contours enclosing 50\% of the total samples, while the thin lines show the contours enclosing 99\% of the total samples.}
        \label{XeXp_hyperposterior_o3_fixedspin_mass_z}
    \end{figure*}

\begin{figure*}[htb]
        \includegraphics[width=2\columnwidth]{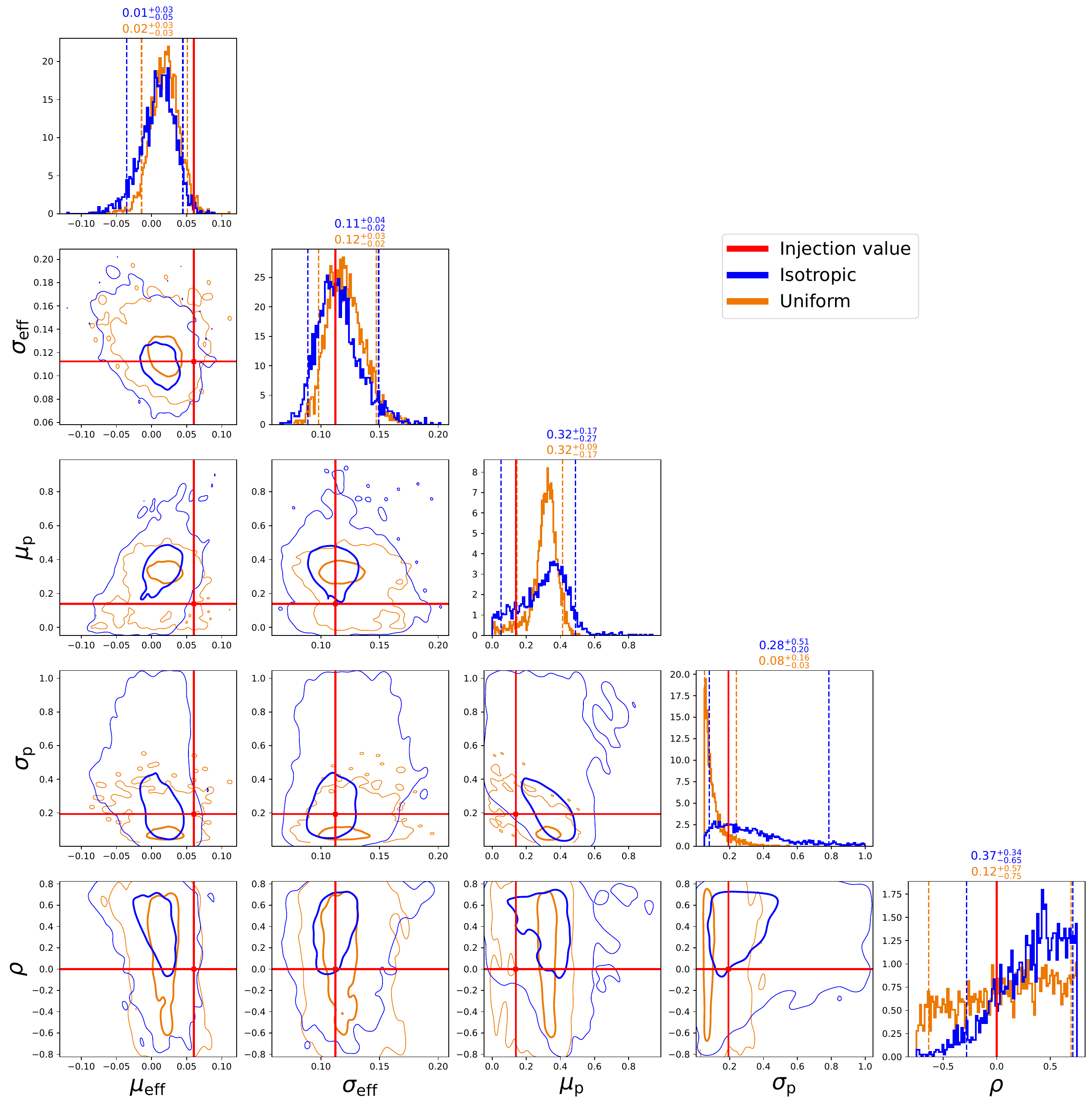}
        \caption{Corner plot for posterior of hyperparameter related to effective spins for GWTC-3-like population case. The population distribution model for the effective spin is a two-dimensional Gaussian distribution, whose hyperparameters consist of two means, two standard deviations, and their correlation. Specifically, $\mu_\mathrm{eff}$ and $\sigma_\mathrm{eff}$ denote the mean and standard deviation of $\Xeff$, while $\mu_\mathrm{p}$ and $\sigma_\mathrm{p}$ denote those of $\Xp$. The parameter $\rho$ represents the correlation between the effective spins. For details, see Sec.~\ref{settings}. The color scheme and plotting conventions are the same as those in Fig.~\ref{XeXp_hyperposterior_o3_fixedspin_mass_z}.}
        \label{XeXp_hyperposterior_o3_fixedspin_spin}
    \end{figure*}

\begin{figure*}[htb]
        \includegraphics[width=2\columnwidth]{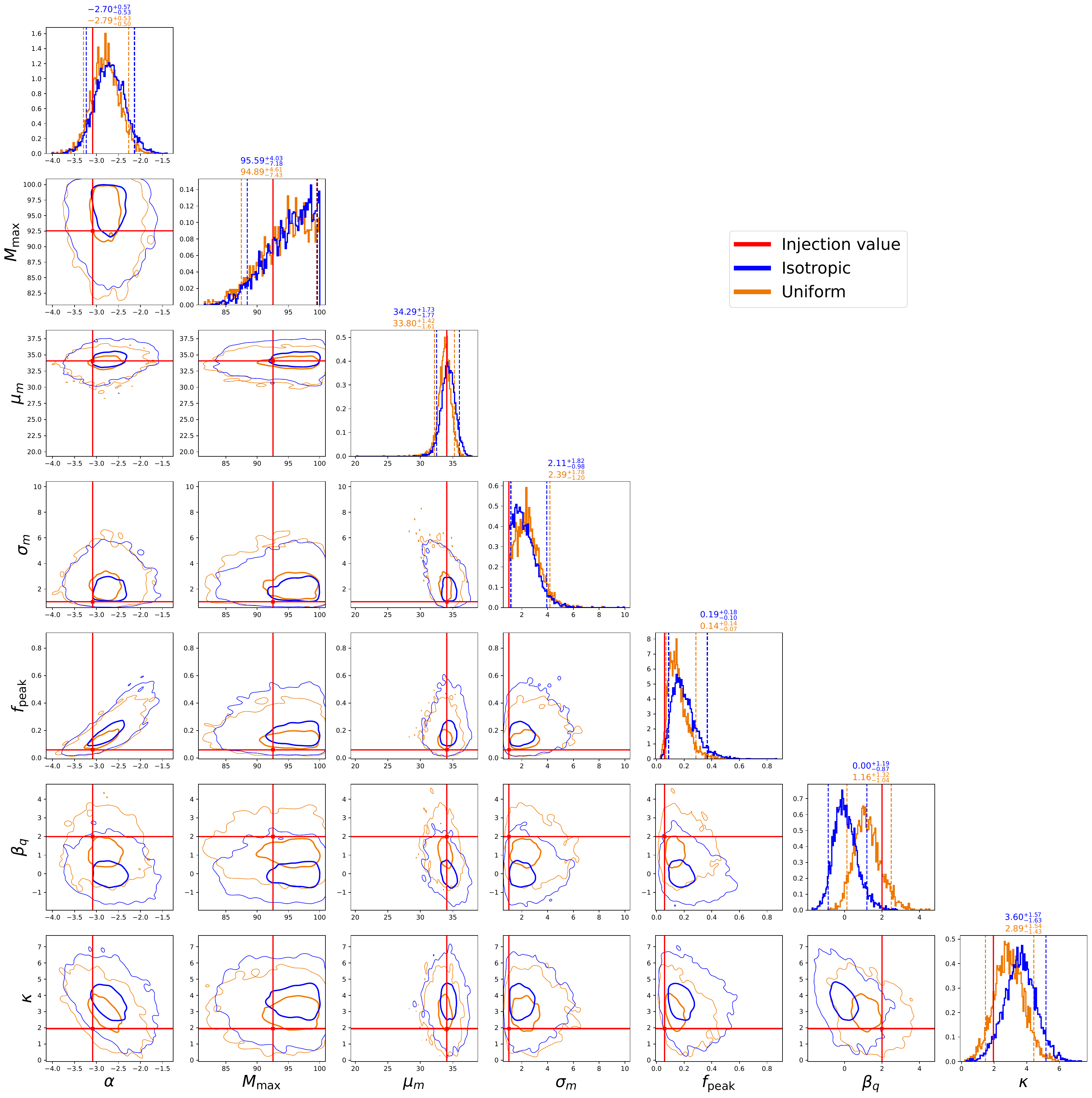}
        \caption{Corner plot for posterior of hyperparameter related to masses and redshift for large-$\Xeff$ population case. The parameters shown in the plot and the plotting conventions are the same as those in Fig.~\ref{XeXp_hyperposterior_o3_fixedspin_mass_z}.}
        \label{XeXp_hyperposterior_largexeff_mass_z}
    \end{figure*}

\begin{figure*}[htb]
        \includegraphics[width=2\columnwidth]{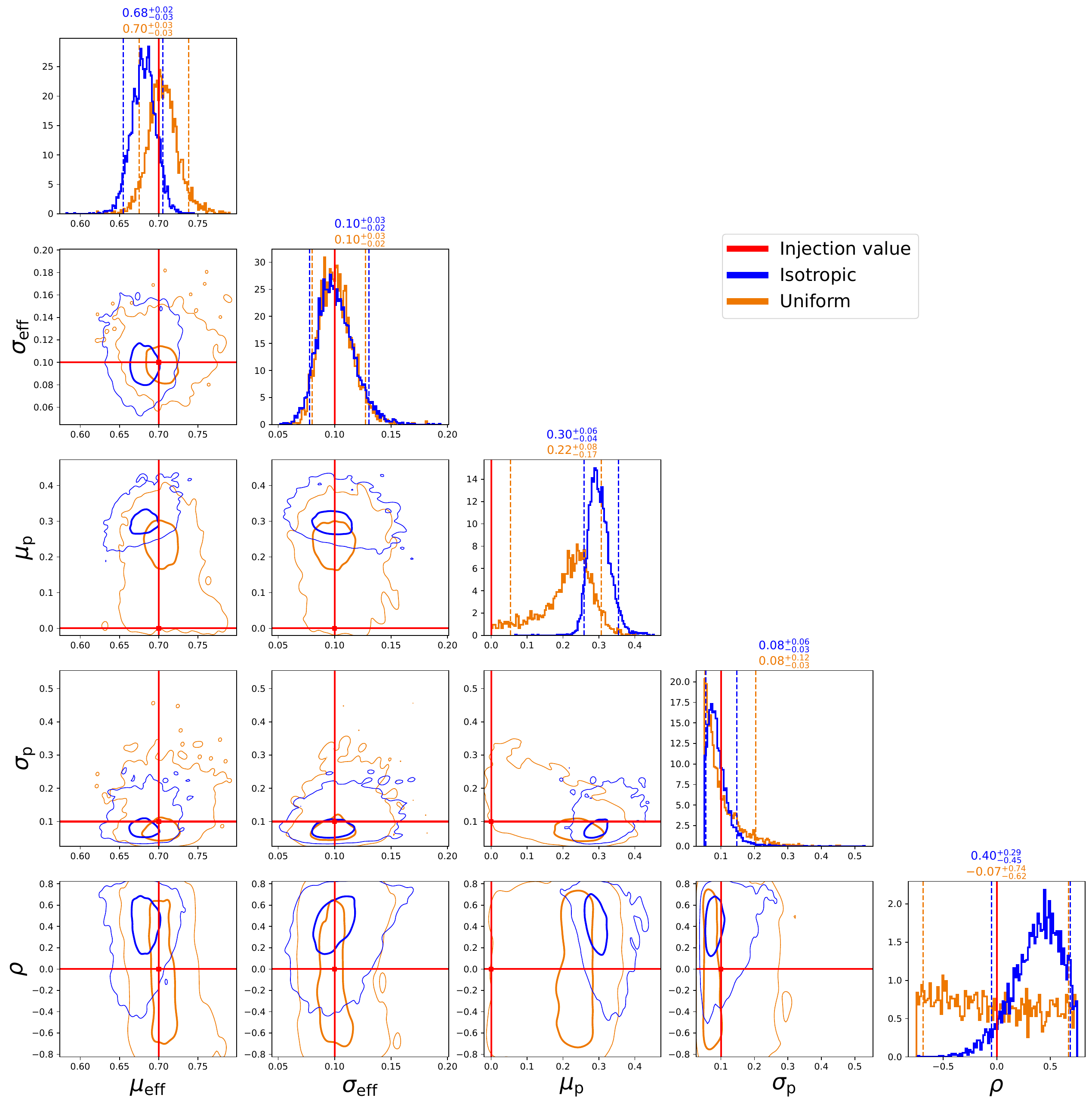}
        \caption{Corner plot for posterior of hyperparameter related to effective spins for large-$\Xeff$ population case. The parameters shown in the plot and the plotting conventions are the same as those in Fig.~\ref{XeXp_hyperposterior_o3_fixedspin_spin}}
        \label{XeXp_hyperposterior_largexeff_spin}
    \end{figure*}

\begin{figure*}[htb]
        \includegraphics[width=2\columnwidth]{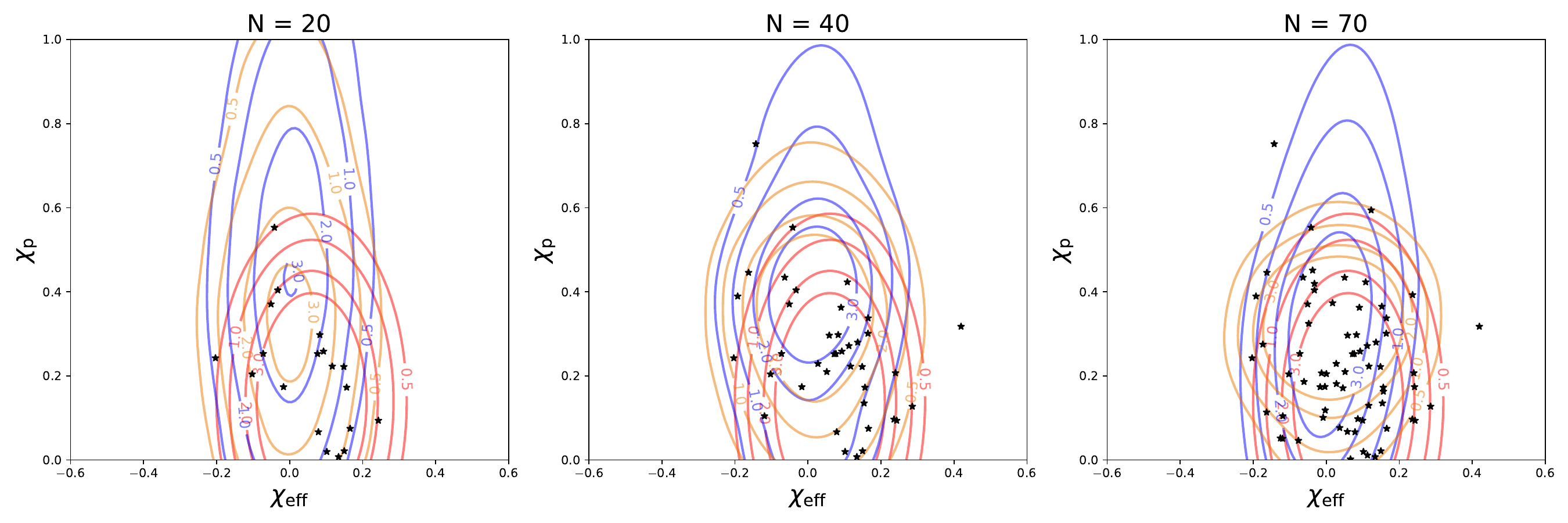}
        
        \includegraphics[width=2\columnwidth]{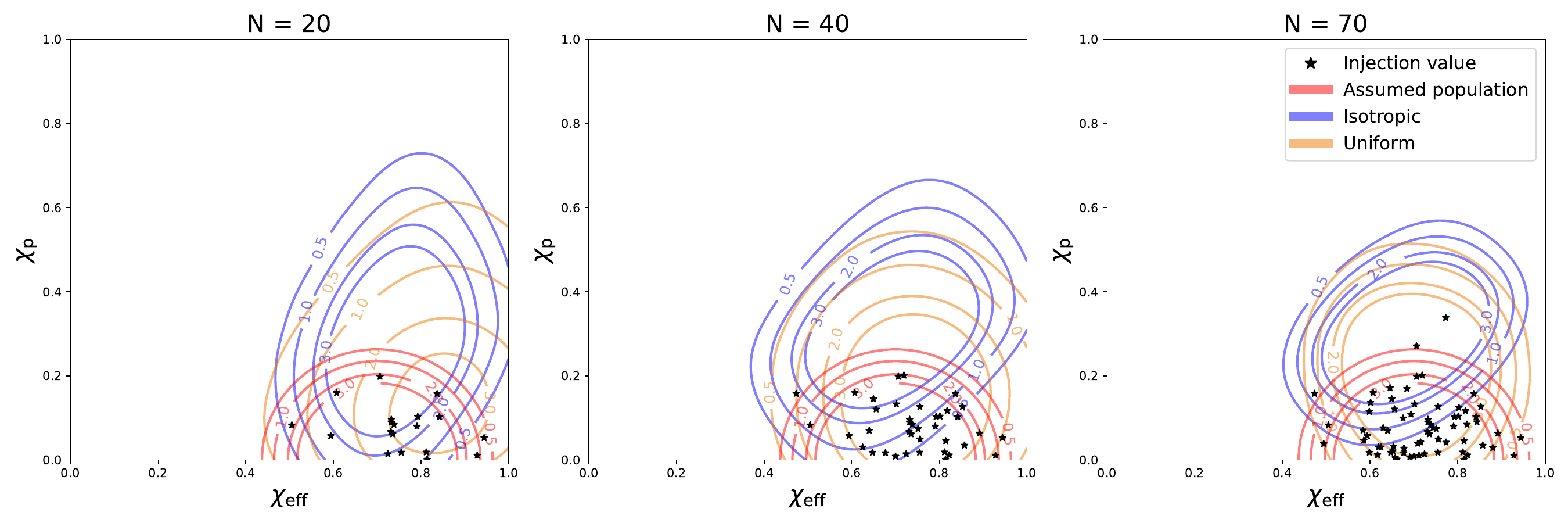}
        \caption{Upper row: inferred population distribution of effective spins for GWTC-3-like injection case. 
        Each sample of the hyperparameters determines a single population distribution. We define the inferred population distribution as the average of these population distributions over the posterior samples of the hyperparameters.
        Red contours show assumed population distribution for each case. The value of $N$ shown at the top of the figure represents the actual number of injections used, and the black star indicates this value. The subsets of $N=20$ and $N=40$ injections consist of the first 20 and 40 injections from the set of $N=70$ injections.
        The contours correspond to $p_\mathrm{pop}(\Xeff, \Xp|\bm{\lambda})$ being 0.5, 1, 2, and 3. 
        Lower row: reconstructed population distribution of effective spins for large-$\Xeff$ injection case. The color scheme is the same as in the upper row.}
        \label{pop_iso_vs_uni_spin}
    \end{figure*}

    Next, we show the results of the hierarchical Bayesian inference.
    Fig.~\ref{XeXp_hyperposterior_o3_fixedspin_mass_z} and Fig.~\ref{XeXp_hyperposterior_o3_fixedspin_spin} show the posterior distributions of the hyperparameters for the GWTC-3–like population case, while Figs.~\ref{XeXp_hyperposterior_largexeff_mass_z} and Fig.~\ref{XeXp_hyperposterior_largexeff_spin} present the results for the large-$\Xeff$ population case. Note that these results are based on hierarchical Bayesian inference using 70 injections.
    Fig.~\ref{pop_iso_vs_uni_spin} shows the reconstructed population distributions of the effective spins. It compares analyses performed with both priors, and also compares the effects of using different numbers of injections in the population inference.



    First, for the GWTC-3–like population case, as shown in Fig.~\ref{XeXp_hyperposterior_o3_fixedspin_mass_z} and Fig.~\ref{XeXp_hyperposterior_o3_fixedspin_spin}, almost all the injected hyperparameter values are included within the 90\% credible intervals of the 1D histogram, regardless of whether the isotropic prior or the $\Xeff$-$\Xp$ uniform prior is adopted. Also, in all the contour plots, the injected values lie within the 99\% contours. This indicates that biases introduced by the isotropic prior on population inference are small for the GWTC-3–like population case.

    For all the hyperparameters except $\mu_\mathrm{p}$, $\sigma_\mathrm{p}$, and $\rho$, the posterior distributions are largely insensitive to the choice of prior. Quantitatively, the Kullback-Leibler (KL) divergences \cite{KL_div} in bits are 0.13, 0.063, 0.42, 0.22, 0.096, 0.20, 0.10, 0.32, and 0.19 for $(\alpha, M_\mathrm{max}, \mu_m, \sigma_m, f_\mathrm{peak}, \beta_q, \kappa, \mu_\mathrm{eff}, \sigma_\mathrm{eff})$, respectively, compared to 0.49, 1.8, and 1.3 for $(\mu_\mathrm{p}, \sigma_\mathrm{p}, \rho)$.

    On the other hand,
    the width of the 90\% credible interval for $\mu_\mathrm{p}$ decreases from 0.44 to 0.26 when the $\Xeff$-$\Xp$ uniform prior is adopted, and a prominent peak emerges around $\mu_\mathrm{p} \approx 0.3$.
    For $\sigma_\mathrm{p}$, adopting the $\Xeff$-$\Xp$ uniform prior produces a sharp peak at $\sigma_\mathrm{p} \lesssim 0.1$. As shown in the 2D correlation plot of $\mu_\mathrm{p}$ and $\sigma_\mathrm{p}$, these parameters are anti-correlated, and this sharp peak is therefore associated with the peak at $\mu_\mathrm{p} \approx 0.3$. 
    
\begin{figure}[htb]
        \includegraphics[width=0.8\columnwidth]{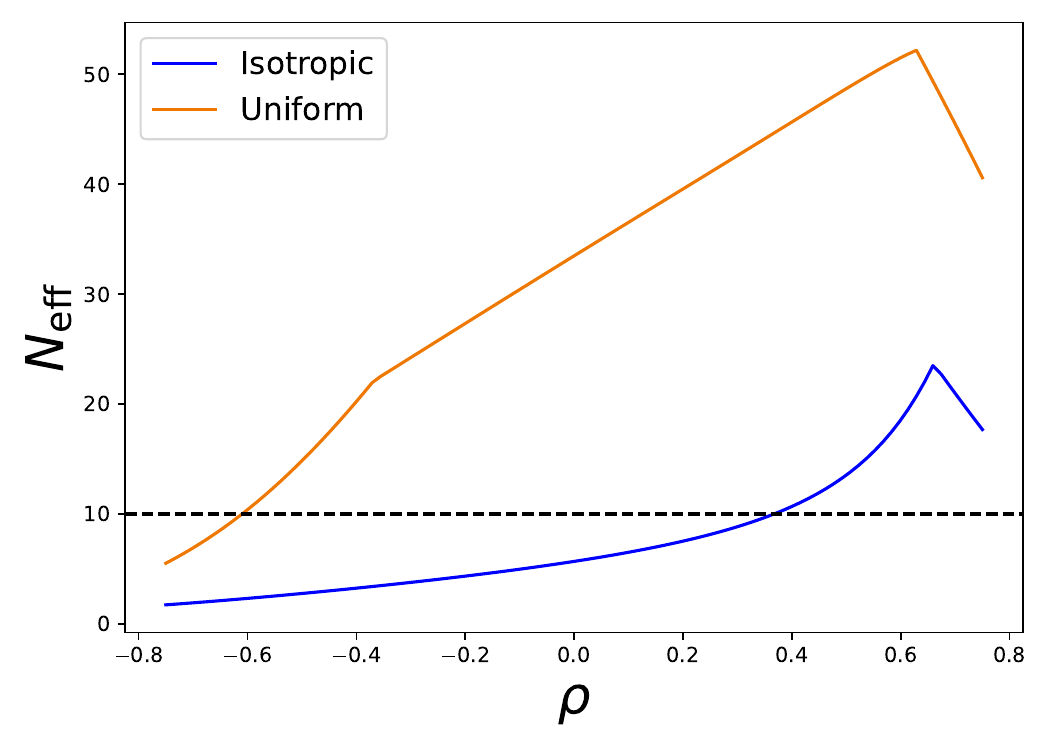}
        \caption{This figure shows $N_\mathrm{eff}$ as a function of $\rho$ in population inference for the GWTC-3–like population. All the hyperparameters except $\rho$ are fixed to their injected values, and $N_\mathrm{eff}$ is computed for the Monte Carlo integration of Eq.~\eqref{hyperposterior_numerical} while varying $\rho$. The blue line corresponds to the isotropic prior, while the orange line corresponds to the $\Xeff$–$\Xp$ uniform prior. The black dashed line indicates the $N_\mathrm{eff}$ threshold.}
        \label{rho_neff}
    \end{figure}

    For the correlation parameter $\rho$ between $\Xeff$ and $\Xp$, the analysis with the isotropic prior shows a mild tendency toward a positive correlation. 
    This behavior is likely a consequence of the structure of the isotropic prior. As shown in the 2D prior distribution in Fig.~\ref{prior_amp_theta}, the region of low prior probability extends to larger $\Xp$ as $|\Xeff|$ increases. Because the assumed population has the positive $\mu_\mathrm{eff}$, the posterior is biased toward larger $\Xp$ for larger $\Xeff$. It artificially produces positive correlation between the effective spins in posterior samples, and reduces $N_\mathrm{eff}$ for population models with negative $\rho$. Fig.~\ref{rho_neff} shows $N_\mathrm{eff}$ as a function of $\rho$, with all other hyperparameters fixed to their injected values. With the isotropic prior, $N_\mathrm{eff}$ drops below the threshold for $\rho \lesssim 0.4$, suppressing the posterior probability in that region of parameter space.
    
    Furthermore, the upper panels of Fig.~\ref{pop_iso_vs_uni_spin} show contours of inferred population distribution of effective spins for the GWTC-3-like population case. From a single posterior sample of hyperparameter $\bm{\lambda}_0$, one obtains a single population distribution of BBH parameters $\bm{\theta}$, denoted as $p_\mathrm{pop}(\bm{\theta}|\bm{\lambda}_0)$. We define the inferred population distribution as the distribution obtained by averaging $p_\mathrm{pop}(\bm{\theta}|\bm{\lambda})$ over the posterior samples of the hyperparameters:
    \begin{align}
        p_\mathrm{pop,inferred}(\bm{\theta})\coloneqq \left\langle p_\mathrm{pop}(\bm{\theta}|\bm{\lambda}) \right\rangle_{\bm{\lambda}\sim p(\bm{\lambda}|D)}.
    \end{align}
    The number $N$ shown at the top of each panel indicates the number of injections used in each result. As can be seen, for both choices of prior, the inferred population distributions tend to converge toward the assumed population as the number of injections increases.
    However, the mean value of $\Xp$ is slightly overestimated when using the isotropic prior. As discussed above, this is because, in the case of the isotropic prior, the posterior distribution of the mean of $\Xp$ has its peak shifted slightly toward larger values.

    On the other hand, we observe significant biases from the isotropic spin prior. As can be seen in Fig.~\ref{XeXp_hyperposterior_largexeff_mass_z}, $\beta_q$ is underestimated when the isotropic prior is employed. As discussed in Sec.~\ref{result_individual_injections}, the mass ratios are underestimated for this prior choice, and these systematic offsets propagate into biased population-level inferences.
    We also observe slight improvements in the estimation of $\kappa$ when the $\Xeff$-$\Xp$ uniform prior is employed. 

    
    We also find that, as shown in Fig.~\ref{XeXp_hyperposterior_largexeff_spin}, significant differences between the two prior choices appear in the inferred spin population. 
    These differences are most pronounced in $\mu_\mathrm{eff}$ and $\mu_\mathrm{p}$: when the isotropic prior is used, $\mu_\mathrm{eff}$ is underestimated, while $\mu_\mathrm{p}$ is overestimated.
    This behavior is directly reflected by the fact, as discussed in Sec.~\ref{result_individual_injections}, that at the level of individual-event analyses, $\Xeff$ tends to be underestimated while $\Xp$ is overestimated. As a consequence, under the isotropic prior, the true population distribution lies too far from the individual-event posterior distributions, causing the reweighting performed in hierarchical Bayesian inference to become ineffective. In practice, this appears as $N_\mathrm{eff}$, falls below the threshold for hyperparameter samples close to the true population. Indeed, for the assumed population, the $N_\mathrm{eff}$ of the hyperparameter samples is 1.26 when using the isotropic prior, whereas it increases to 48.3 when adopting the $\Xeff$-$\Xp$ uniform prior.
    In addition, $\rho$ is overestimated with the isotropic prior due to the threshold on $N_{\mathrm{eff}}$, as is the case for the GWTC-3-like population.


    Finally, the lower panels of Fig.~\ref{pop_iso_vs_uni_spin} show 
    inferred population distributions of effective spins for the large-$\Xeff$ population case. From this figure, we find that when adopting the $\Xeff$-$\Xp$ uniform prior, the inferred population becomes increasingly consistent with the assumed population as the number of injections used in the analysis increases. In contrast, when using the isotropic prior, significant biases in the inferred population are observed even as the number of injections increases, particularly for $\Xp$. Since this effect is clearly driven by the choice of prior, as discussed above, these results demonstrate that careful consideration of the prior used for individual events is essential in population inference.

\section{Conclusion}
\label{conclusion}
    
    In this study, we investigate the impact of the assumed spin prior, focusing on both individual parameter estimation and population inference.
    We introduce the novel spin prior focusing on $\Xeff$ and $\Xp$, motivated by the fact that these parameters can be measured relatively well through GW observations. Specifically, this prior is uniform over the physically allowed region of the effective spin parameters conditioned on the mass ratio, and therefore assigns equal support to regions that are strongly disfavored by the commonly used isotropic prior, such as those with $\Xp \approx 0$ and large $|\Xeff|$.

    We investigate the impact of the spin prior by analyzing BBH injections from mock populations with both the isotropic and $\Xeff$-$\Xp$ uniform priors. 
    Specifically, we perform simulations for two mock populations: a GWTC-3-like population whose spins lie in a region well supported by the isotropic prior, and a population distributed at large $\Xeff$, where the isotropic prior assigns vanishing probability. 

    Our individual parameter estimation studies show that, for the GWTC-3-like population, both of the priors recover the injected values reasonably well, although the isotropic prior slightly tends to disfavor around $\Xp \approx 0$. Meanwhile, for the large-$\Xeff$ population, we find that individual parameter estimation using the $\Xeff$-$\Xp$ uniform prior correctly recovers the injected values, whereas the isotropic prior exhibits systematic biases, underestimating $\Xeff$ and overestimating $\Xp$. We further find that, accompanying the underestimation of $\Xeff$, the chirp mass and mass ratio are also underestimated in parameter estimation using the isotropic prior for the large-$\Xeff$ population case.
    
    At the population-inference level, the impact of the prior becomes even more pronounced. For the GWTC-3-like population, with either prior, the inferred spin distribution converges toward the true underlying spin distribution as the number of injections increases. However, the $\Xp$ distribution recovered with the isotropic prior is slightly shifted toward larger values relative to the true distribution. 
    In contrast, in the large-$\Xeff$ population case, population inference with the isotropic prior exhibits significant systematic biases. Specifically, the inferred $\Xeff$ distribution is systematically shifted toward smaller values, closer to $\Xeff = 0$, than the true underlying population, while the inferred $\Xp$ distribution is shifted toward larger values. 
    These biases arise from the structure of the isotropic prior on the effective spins.

    Overall, our results demonstrate that the choice of prior plays a critical role in spin inference at both the event and population levels. In particular, population inference related to $\Xp$ is highly sensitive to the analysis setup.
    While the isotropic prior performs adequately for populations that are well aligned with its support, it introduces biases when the true spin distribution favors large aligned components.
    In particular, the presence of such systematic biases can be critical for analyses searching for subpopulations with large $\Xeff$ \cite{large_xeff_subpop, hierchical_xeffxp_subpop}.
    Moreover, such biases do not diminish as the number of events increases. In fact, they become increasingly pronounced. Therefore, the presence of such systematic biases would be critical for future observations, in which the number of detected events is expected to grow substantially. 
    For this reason, careful consideration must be given to the choice of priors for individual events when performing population inference.

    Since this bias can affect not only populations with large $\Xeff$ but also, to a lesser extent, GWTC-3-like populations, previous analyses of BBH populations may also be subject to similar biases.
    An analysis of real observational data from the LVK collaboration using the method presented in this paper, as well as the corresponding publication, is currently in preparation.

\section*{Acknowledgments}

    We are grateful to Tejaswi Venumadhav for useful discussions at the early stage of this work. We are indebted to Thomas Callister for sharing the GWTC-3 population analysis code. We appreciate Daiki Watarai, Hayato Imafuku and Takahiro S. Yamamoto for insightful comments. We also acknowledge Leo Tsukada for carefully reviewing the manuscript and Hui Tong for helpful discussions. 
    K. K. is supported by JST SPRING, Grant Number JPMJSP2108.
    M. I. acknowledges the support from the Royal Society Award ICA\textbackslash R1\textbackslash 231114.
    S. M. acknowledges support from JSPS Grant-in-Aid for Transformative Research Areas (A) No.~23H04891 and No.~23H04893.
    K. H. work was supported by JSPS KAKENHI Grant-in-Aid for Scientific Research JP23H01169, JP23H04900, and JST FOREST Program JPMJFR2136.
    T. K. is supported from JSPS Grant-in-Aid for Scientific Research (C) No.~22K03630, JSPS Grant-in-Aid for Transformative Research Areas (A) 23H04893 and the financial support from the Science Moves Award.

\section*{Data availability}
    The source code used for the analysis is available at GitHub \cite{Xeff_Xp_uniform_prior}. The underlying large numerical data are available from the corresponding author upon request.

    \appendix
    \begin{widetext}
    \section{Calculations of the prior distributions}
    \label{Calculations_of_prior}
    In this study, we newly adopt the prior of the spins that is simultaneously uniform with respect to $\Xeff$ and $\Xp$ for given $q$. The Explicit form of this prior is given in Eq.~\eqref{Xe_Xp_uni_prior}. In this appendix, we describe various calculations associated with this prior.

    First, we derive the alternative form of the prior distribution, Eq.~\eqref{Xe_Xp_uni_prior_alt}. The trick here is the equivalent transformations:
    \begin{align}
        &\qty(|\Xeff|\leq\frac{q}{1+q})\cap\qty(0<\Xp<1)\notag\\
        \iff&\qty(|\Xeff|\leq\frac{q}{1+q})\cap\qty(0<\Xp<1)\cap\qty(|\Xeff|<\frac{\sqrt{1-\Xp^2}+q}{1+q}),\\
        &\qty(\frac{q}{1+q}<|\Xeff|<1)\cap\qty(0<\Xp<\sqrt{1-\qty((1+q)|\Xeff|-q)^2})\notag\\
        \iff&\qty(\frac{q}{1+q}<|\Xeff|<1)\cap\qty(0<\Xp)\cap\qty(\Xp^2<1-\qty((1+q)|\Xeff|-q)^2)\notag\\
        \iff&\qty(\frac{q}{1+q}<|\Xeff|<1)\cap\qty(0<\Xp<1)\cap\qty((1+q)|\Xeff|-q<\sqrt{1-\Xp^2})\notag\\
        \iff&\qty(\frac{q}{1+q}<|\Xeff|)\cap\qty(0<\Xp<1)\cap\qty(\abs{\Xeff}<\frac{\sqrt{1-\Xp^2}+q}{1+q}).
    \end{align}
    Therefore, we have, in light of Eq.~\eqref{X_p_max}, the following relation holds:
    \begin{align}
        \qty(\abs{\Xeff}<1)\cap\qty(0<\Xp<\chi_\text{p\_max}(q,\Xeff))\iff\qty(0<\Xp<1)\cap\qty(\abs{\Xeff}<\frac{\sqrt{1-\Xp^2}+q}{1+q}),
    \end{align}
    which provides Eq.~\eqref{Xe_Xp_uni_prior_alt}.
    
    Next, we derive the normalization factor $N_\chi(q)$.
    Integrating both sides of Eq.~\eqref{Xe_Xp_uni_prior_alt} over the entire parameter space reads
    \begin{align}
        1
        &= N_\chi(q)\int \mathrm{d}\Xeff\int\mathrm{d}\Xp \Theta\qty(0<\Xp< 1)\Theta\qty(\abs{\Xeff}<\frac{\sqrt{1-\Xp^2}+q}{1+q})\\
        \therefore \frac{1}{N_\chi(q)} &= \int \mathrm{d}\Xeff\int\mathrm{d}\Xp \Theta\qty(0<\Xp< 1)\Theta\qty(\abs{\Xeff}<\frac{\sqrt{1-\Xp^2}+q}{1+q})\notag\\
        &= 2\int_0^1\mathrm{d}\Xp \frac{\sqrt{1-\Xp^2}+q}{1+q}=\frac{2}{1+q}\qty(\frac{\pi}{4}+q).
    \end{align}
    Therefore, the normalization factor is shown to be
    \begin{align}
        N_\chi(q) = \frac{2+2q}{\pi +4q}.
    \end{align}
    The derivations of marginal distributions for either $\Xeff$ or $\Xp$ are trivial, so we do not describe them here.

    Next, we describe in detail the prior settings for the spin parameters other than the effective spins in the prior implemented in this study.
    
    As described in Sec.~\ref{uniform_xexpu_prior_implementation}, in this study, we adopt the prior that is defined such that, given $\Xeff$, $\Xp$, and $q$, the marginalized distributions of $\XIz$ and $\XIIz$ are uniform over their allowed parameter region, and given $\Xeff$, $\Xp$, $q$, $\XIz$ and $\XIIz$, the conditional distribution of $\XIp$ and $\XIIp$ is proportional to $\XIp\times\XIIp$ over their allowed parameter region.
    For instance, considering only the $z$-components, once $\Xeff$ and $q$ is specified, the permissible values of $\XIz$ and $\XIIz$ lie solely on a line defined by $\XIz= -q\XIIz+(1+q)\Xeff$. 
    For convenience, we consider the spins in a cylindrical coordinate system defined by $\XIz, \XIIz, \XIp, \XIIp, \phi_{JL}, \phi_{12}$. Here, $\phi_{JL}$ and $\phi_{12}$ represent the azimuthal angles of the spins. Detailed definitions can be found in \cite{bilby_cbc}.
    
    Mathematically, the conditional prior distribution for $\XIz, \XIIz$ is set to
    \begin{align}
    &\pi_\mathrm{PE}(\XIz, \XIIz|\Xeff, \Xp, q)\mathrm{d}\XIz\mathrm{d}\XIIz\notag\\
    &= N_\mathrm{z} \delta \left(\Xeff - \Xeff(\XIz, \XIIz, q)\right)
    \Theta\left(-1< \XIz < 1\right)\Theta\left(-1< \XIIz < 1\right)\notag\\
    &\quad\times\Theta\left[\qty(\Xp<\sqrt{1-\XIz^2}) \  \cup \  \qty(\Xp< \frac{3+4q}{4+3q}q\sqrt{1-\XIIz^2})\right]\mathrm{d}\XIz\mathrm{d}\XIIz,
    \label{p_X1z_X2z}
    \end{align}
    where $N_\mathrm{z}$ is a normalization factor, $\delta$ denotes the delta function, and $\Theta$ is the Heaviside step function that takes the value 1 only when the specified conditions are satisfied, and 0 otherwise. It should also be noted that $\Xeff$ and $\Xeff(\XIz, \XIIz, q)$ are different. $\Xeff$ is given as an argument, but $\Xeff(\XIz, \XIIz, q)$ is the value determined by $\XIz$, $\XIIz$, and $q$. In other words, it is simply a replacement by $(\XIz + q \XIIz) / (1+q)$. We have also omitted $\Theta$-functions corresponding to conditions that are obviously satisfied by the given values, such as $\Theta(-1 < \Xeff < 1)$. The same applies below.
    
    The conditional prior distribution for $\XIp, \XIIp$ is set to
    \begin{align}
    &\pi_\mathrm{PE}(\XIp, \XIIp|\Xp, q, \XIz, \XIIz)\mathrm{d}\XIp\mathrm{d}\XIIp\notag\\
    &= N_\mathrm{p}  \XIp\XIIp\delta \left(\Xp - \Xp(\XIp, \XIIp, q)\right)\Theta\left(0< \XIp < \sqrt{1-\XIz^2}\right)\Theta\left(0< \XIIp < \sqrt{1-\XIIz^2}\right)\mathrm{d}\XIp\mathrm{d}\XIIp,
    \label{p_X1p_X2p}
    \end{align}
    where $N_\mathrm{p}$ is a normalization factor and the quantity $\Xp(\XIp, \XIIp, q)$ is also obtained by replacing $\mathrm{max}\left(\XIp, ((3+4q)/(4+3q))q\XIIp\right)$.
    
    The priors for the spin azimuthal angles are a uniform distribution ranging from 0 to $2\pi$:
    \begin{align}
        \pi_\mathrm{PE}(\phi_{JL})\mathrm{d}\phi_{JL}&=\frac{1}{2\pi}\Theta\left(0< \phi_{JL} < 2\pi\right)\mathrm{d}\phi_{JL}\label{p_phi_jl}\\
        \pi_\mathrm{PE}(\phi_{12})\mathrm{d}\phi_{12}&=\frac{1}{2\pi}\Theta\left(0< \phi_{12} < 2\pi\right)\mathrm{d}\phi_{12}\label{p_phi_12}.
    \end{align}

    Consequently, the overall prior for the spins without azimuthal angles can be expressed using Eq.~\eqref{Xe_Xp_uni_prior} and Eqs.~\eqref{p_X1z_X2z}–\eqref{p_phi_12} as
    \begin{align}
        &\pi_\mathrm{PE}(\Xeff, \Xp, \XIz, \XIIz, \XIp, \XIIp, \phi_{JL}, \phi_{12}|q)\notag\\
        &=\pi_\mathrm{PE}\left(\Xeff, \Xp|q\right)\pi_\mathrm{PE}(\XIz, \XIIz|\Xeff, \Xp, q)\pi_\mathrm{PE}(\XIp, \XIIp|\Xp, q, \XIz, \XIIz)\pi_\mathrm{PE}(\phi_{JL})\pi_\mathrm{PE}(\phi_{12}).
    \end{align}
    At first glance, the left-hand side appears to contain two additional spin degrees of freedom. However, since the right-hand side includes two delta functions within the expression, the number of independent degrees of freedom is indeed six, and thus there is no inconsistency.

    The next issue is how to generate samples that follow this probability density function. Since the sampling for $\Xeff$, $\Xp$, $\phi_{JL}$, and $\phi_{12}$ is straightforward, we do not discuss it in detail here.

    First, for the sampling of $\XIz$ and $\XIIz$, the corresponding function can be regarded as a product of a marginal distribution and a conditional distribution:
    \begin{align}
        \pi_\mathrm{PE}(\XIz, \XIIz|\Xeff, \Xp, q) = \pi_\mathrm{PE}(\XIz|\Xeff, \Xp, q)\pi_\mathrm{PE}(\XIIz|\Xeff, \Xp, q, \XIz),
        \label{p_X1z_X2z_div}
    \end{align}
    where $\pi_\mathrm{PE}(\XIz|\Xeff, \Xp, q)$ denotes the marginal distribution, defined as
    \begin{align}
        \pi_\mathrm{PE}(\XIz|\Xeff, \Xp, q) \coloneqq \int \mathrm{d} \XIIz \pi_\mathrm{PE}(\XIz, \XIIz|\Xeff, \Xp, q),
        \label{p_X1z}
    \end{align}
    and the conditional distribution is written as
    \begin{align}
        \pi_\mathrm{PE}(\XIIz|\Xeff, \Xp, q, \XIz) \coloneqq \frac{\pi_\mathrm{PE}(\XIz, \XIIz|\Xeff, \Xp, q)}{\pi_\mathrm{PE}(\XIz|\Xeff, \Xp, q)}.
        \label{p_X2z}
    \end{align}
    In practice, the algorithm first generates a sample according to Eq.~\eqref{p_X1z}, and then uses this value to generate a sample according to Eq.~\eqref{p_X2z}. This procedure is equivalent to obtaining a sample from the joint probability distribution given in Eq.~\eqref{p_X1z_X2z_div}.
    Thus, the problem reduces to how to generate samples from each of the one-dimensional probability distributions.
    If the inverse of the CDF for each of these distributions can be obtained, samples following the corresponding probability densities can be generated using the inverse transform method.
    We now proceed to compute the CDF of Eq.~\eqref{p_X1z}. First, by evaluating the integral, we obtain
    \begin{align}
        &\pi_\mathrm{PE}(\XIz|\Xeff, \Xp, q)\notag\\ 
        &\propto
         \Theta\qty(\abs{\XIz}<1)\Theta\qty(\abs{(1+q)\Xeff-\XIz}<q)
         \Theta\left[\qty(\Xp<\sqrt{1-\XIz^2}) \  \cup \  
         \qty(\nu\Xp<\sqrt{q^2-\qty((1+q)\Xeff-\XIz)^2})\right]
    \end{align}
    where we define $\nu \coloneqq (4+3q)/(3+4q)$. This distribution is a uniform distribution with boundaries dependent on $\Xeff,\Xp,q$. The conditions here can be expressed by the union of two conditions:
    \begin{align}
        &\qty(\abs{\XIz}<1)\cap\qty(\abs{(1+q)\Xeff-\XIz}<q)\cap\qty[\qty(\Xp<\sqrt{1-\XIz^2}) \  \cup \  
         \qty(\nu\Xp<\sqrt{q^2-\qty((1+q)\Xeff-\XIz)^2})]\notag\\
         \iff&\qty[\qty(\abs{\XIz}<\sqrt{1-\Xp^2})\cap\qty(\abs{(1+q)\Xeff-\XIz}<q)]\notag\\&\cup\qty[\qty(\abs{\XIz}<1)\cap\qty(q^2-\nu^2\Xp^2>\qty((1+q)\Xeff-\XIz)^2)]
    \end{align}
    Therefore, the domain where this distribution has non-zero support is either a single interval or a union of two intervals, because each sub-condition indicates an interval. Specifically, if $q<\nu\Xp$, that is, when $\Xp$ is so large that even $\XIIp=1$ cannot realize $\Xp$, the second condition is never satisfied and therefore the domain of this marginal distribution becomes a single interval:
    \begin{align}
        \qty(\max\qty(-\sqrt{1-\Xp^2},-q+(1+q)\Xeff),\min\qty(\sqrt{1-\Xp^2},q+(1+q)\Xeff)).
    \end{align}
    If $q\geq\nu\Xp$, on the other hand, the second condition contributes to the domain and makes the boundary complicated. For this case, the region where $\XIz$ can physically take can be described by
    \begin{align}
        &\qty(\max\qty(-\sqrt{1-\Xp^2},-q+(1+q)\Xeff),\min\qty(\sqrt{1-\Xp^2},q+(1+q)\Xeff))\notag\\
        &\cup\qty(\max\qty(-1,-\sqrt{q^2-\nu^2\Xp^2}+(1+q)\Xeff),\min\qty(1,\sqrt{q^2-\nu^2\Xp^2}+(1+q)\Xeff)).
    \end{align}
    This becomes single interval if $(1+q)\abs{\Xeff}\leq\sqrt{1-\Xp^2}+\sqrt{q^2-\nu^2\Xp^2}$ is satisfied. otherwise, the allowed domain for $\XIz$ consists of two intervals. 
    In either case, since the distribution is uniform over the resulting range, both the CDF and its inverse are straightforward.

    Next, for $\XIIz$, once $q,\Xeff,$ and $\XIz$ are given, its value is uniquely determined.  
    Therefore, the probability density function is expressed as
    \begin{align}
        \pi_\mathrm{PE}(\XIIz | \Xeff, \Xp, q, \XIz) &\propto \delta \left(\Xeff - \Xeff(\XIz, \XIIz, q)\right)\notag\\
        &=\delta \left(\Xeff - \frac{\XIz + q\XIIz}{1+q}\right),
    \end{align}
    and the CDF of this distribution is 
    \begin{align}
        \mathrm{CDF}(\XIIz | \Xeff, \Xp, q, \XIz) = \Theta \left(\XIIz > \frac{(1+q)\Xeff - \XIz}{q}\right).
    \end{align}

    Next, for $\XIp$ and $\XIIp$, similar to $\XIz$ and $\XIIz$, the distribution is decomposed into the marginalized distribution of $\XIp$ and the conditional distribution of $\XIIp$, and sampling is performed in the order $\XIp \rightarrow \XIIp$.

    First, regarding the marginalized distribution of $\XIp$, it can be written as
    \begin{align}
        \pi_\mathrm{PE}(\XIp|\Xp, q, \XIz, \XIIz) &\coloneqq \int \mathrm{d} \XIIp \pi_\mathrm{PE}(\XIp, \XIIp|\Xp, q, \XIz, \XIIz)\notag\\
        &= N_\mathrm{p}  \XIp \Theta\left(0< \XIp < \sqrt{1-\XIz^2}\right)\notag\\
        &\quad\times\int\mathrm{d}\XIIp\XIIp\delta \left[\Xp - \mathrm{max}\left(\XIp, \frac{q\XIIp}{\nu}\right)\right]\Theta\left(0< \XIIp < \sqrt{1-\XIIz^2}\right).
        \label{p_X1p}
    \end{align}
    Here, the integral in this expression can be separated into the following two terms:
    \begin{align}
        &\int\mathrm{d}\XIIp\XIIp\delta \left[\Xp - \mathrm{max}\left(\XIp, \frac{q\XIIp}{\nu}\right)\right]\Theta\left(0< \XIIp < \sqrt{1-\XIIz^2}\right)\notag\\
        &=\int\mathrm{d}\XIIp\XIIp\delta \left(\Xp - \XIp\right)\Theta\left(\XIp > \frac{q\XIIp}{\nu}\right)\Theta\left(0< \XIIp < \sqrt{1-\XIIz^2}\right) \notag\\
        &\quad+ \int\mathrm{d}\XIIp\XIIp\delta \left(\Xp - \frac{q\XIIp}{\nu}\right)\Theta\left(\XIp < \frac{q\XIIp}{\nu}\right)\Theta\left(0< \XIIp < \sqrt{1-\XIIz^2}\right)\notag\\
        &=\frac{1}{2}\delta \left(\Xp - \XIp\right)\left[\min\left(\frac{\nu\XIp}{q}, \sqrt{1-\XIIz^2}\right)\right]^2
        + \left(\frac{\nu}{q}\right)^2\Xp\Theta\left(\XIp < \Xp < \frac{q}{\nu}\sqrt{1-\XIIz^2}\right).
    \end{align}
    Here, the integral of $\pi_\mathrm{PE}(\XIIz | \Xeff, \Xp, q, \XIz)$ over $\XIIz$ is, of course, equal to 1.  
    We then compute the contributions of each term, denoted $N_1$ and $N_2$, respectively. 
    Specifically,  
    \begin{align}
        N_1 \coloneqq& \int \mathrm{d}\XIp N_\mathrm{p} \XIp \Theta\left(0< \XIp < \sqrt{1-\XIz^2}\right) \frac{1}{2}\delta \left(\Xp - \XIp\right)\left[\min\left(\frac{\nu\XIp}{q}, \sqrt{1-\XIIz^2}\right)\right]^2\notag\\
        =&\frac{N_\mathrm{p}}{2}\Xp\left[\min\left(\frac{\nu\Xp}{q}, \sqrt{1-\XIIz^2}\right)\right]^2\Theta\left(0< \Xp < \sqrt{1-\XIz^2}\right),\\
        N_2 \coloneqq& \int \mathrm{d}\XIp N_\mathrm{p} \XIp \Theta\left(0< \XIp < \sqrt{1-\XIz^2}\right) \left(\frac{\nu}{q}\right)^2\Xp\Theta\left(\XIp < \Xp < \frac{q}{\nu}\sqrt{1-\XIIz^2}\right)\notag\\
        =&\frac{N_\mathrm{p}}{2} \left(\frac{\nu}{q}\right)^2\Xp \left[\min\left(\sqrt{1-\XIz^2}, \Xp\right)\right]^2\Theta\left(\Xp < \frac{q}{\nu}\sqrt{1-\XIIz^2}\right).
    \end{align}
    Since $N_1 + N_2 = 1$, the overall normalization factor is
    \begin{align}
        \frac{1}{N_\mathrm{p}}
        &= \frac{1}{2}\Xp\left[\min\left(\frac{\nu\Xp}{q}, \sqrt{1-\XIIz^2}\right)\right]^2\Theta\left(0< \Xp < \sqrt{1-\XIz^2}\right)\notag\\
        &\quad+ \frac{1}{2} \left(\frac{\nu}{q}\right)^2\Xp \left[\min\left(\sqrt{1-\XIz^2}, \Xp\right)\right]^2\Theta\left(\Xp < \frac{q}{\nu}\sqrt{1-\XIIz^2}\right).
    \end{align}
    Therefore, one can sample from the first term with probability $N_1$, and from the second term with probability $N_2 = 1 - N_1$. Specifically, since the first term contains $\delta \left(\Xp - \XIp\right)$, in this case one should always return just $\Xp$. On the other hand, for the second term:
    \begin{align}
         N_\mathrm{p} \left(\frac{\nu}{q}\right)^2\Xp \XIp 
         \Theta\left(0< \XIp < \sqrt{1-\XIz^2}\right)\Theta\left(\XIp < \Xp < \frac{q}{\nu}\sqrt{1-\XIIz^2}\right),
    \end{align}
    the probability density function is linear probability density function. The derivation of its CDF and inverse function is omitted.

    Finally, for the sampling of $\XIIp$, the behavior depends on whether $\XIp$ corresponds to the $N_1$ case or the $N_2$ case.  
    In the former case, $\XIp$ is fixed at $\Xp$, so that
    \begin{align}
        \pi_\mathrm{PE}(\XIIp | \Xp, q, \XIz, \XIIz, \XIp) \coloneqq \frac{\pi_\mathrm{PE}(\XIp, \XIIp | \Xp, q, \XIz, \XIIz)}{\pi_\mathrm{PE}(\XIp | \Xp, q, \XIz, \XIIz)},
    \end{align}
    which yields
    \begin{align}
        \pi_\mathrm{PE}(\XIIp | \Xp, q, \XIz, \XIIz) \propto \XIIp \Theta\left(\XIIp < \frac{\Xp \nu}{q}\right) \Theta\left(0 < \XIIp < \sqrt{1-\XIIz^2}\right).
    \end{align}
    Since this is a linear probability density function, sampling from it is straightforward.
    In the latter case, the distribution is given by
    \begin{align}
        \pi_\mathrm{PE}(\XIIp \mid \Xp, q, \XIz, \XIIz, \XIp) \propto \delta \left(\Xp - \frac{q \XIIp}{\nu}\right).
    \end{align}
    Accordingly, $\XIIp$ is fixed at $\XIIp = \Xp \nu/q$. 
    
    Thus, the spin parameters in Cartesian coordinates are drawn conditioned on the specified values of $\Xeff$, $\Xp$, and $q$.
    Fig.~\ref{xexpuni_prior_other} shows the distributions for $\XIz,\XIIz,\XIp$ and $\XIIp$ in our $\Xeff$-$\Xp$ uniform prior distribution.

    \begin{figure}[htb]
        \includegraphics[width=1\columnwidth]{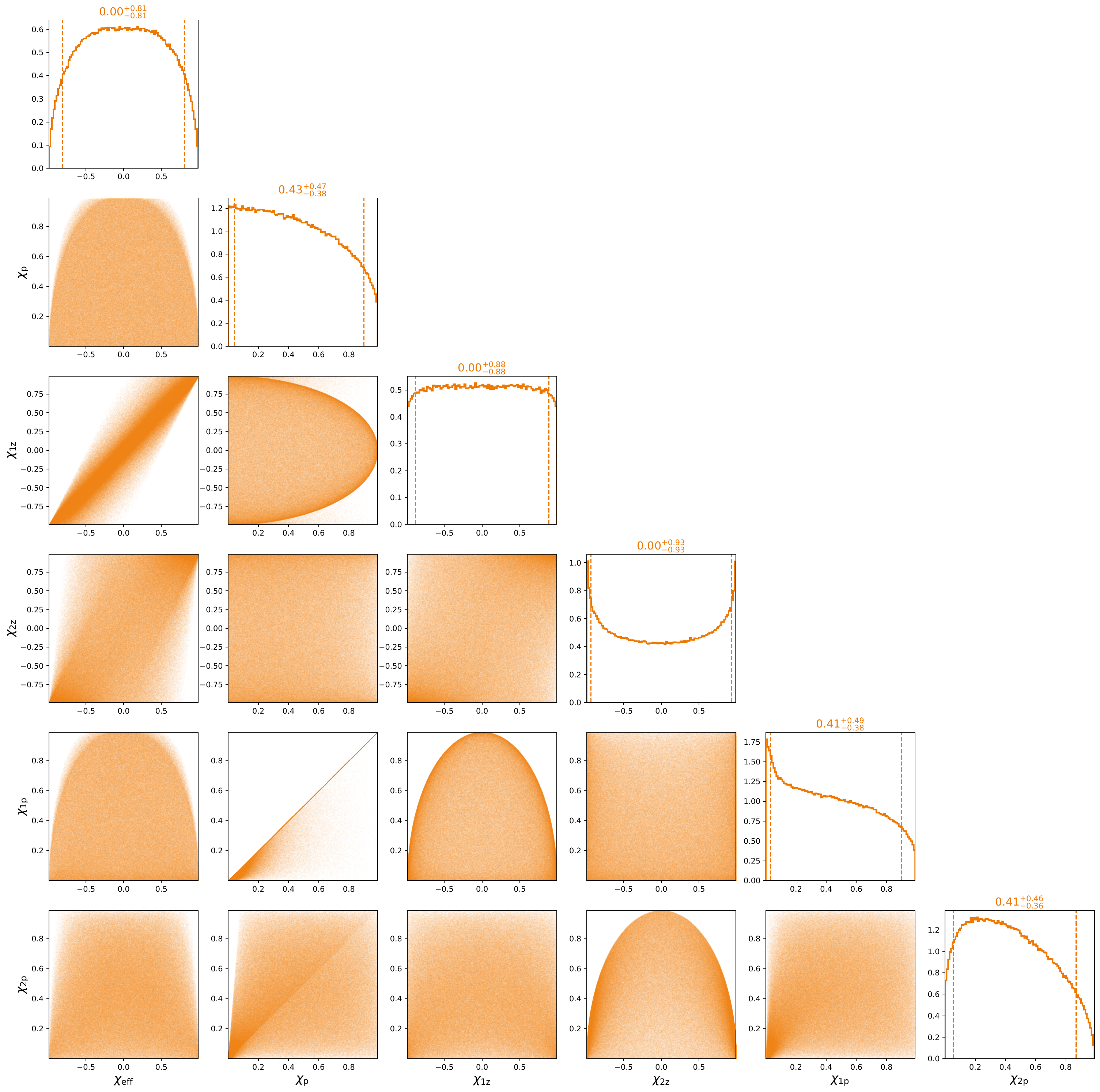}
        \caption{Corner plot of the spin parameters under the $\Xeff$-$\Xp$ uniform prior in our implementation. The plot is generated from samples drawn according to that prior distribution. Here, the prior on $q$ is chosen such that $m_1$ and $m_2$ are uniformly distributed. The dashed lines show 90\% confidence lines. 
        }
        \label{xexpuni_prior_other}
    \end{figure}

    \section{Robustness Analyses}
    \label{Robustness_Analyses}

    \subsection{Sensitivity to the choice of the residual spin prior}

    While the marginal prior on $(\Xeff,\Xp)$ is fixed by construction, different conditional distributions for the residual spin degrees of freedom can lead to the same $(\Xeff,\Xp)$ prior. To assess the impact of this ambiguity, we perform a sensitivity test using an alternative completion of the residual spin prior.
    Specifically, we consider an alternative conditional prior for $(\XIp,\XIIp)$ obtained by removing the $\times\XIp\XIIp$ factor from Eq.~\eqref{p_X1p_X2p}. 
    This also preserves the marginal prior on $(\Xeff,\Xp)$ while providing a different completion of the six-dimensional spin prior.


    We reweight the posterior samples obtained with the $\Xeff$-$\Xp$ uniform prior to the alternative prior and compare the resulting posterior distributions.
    To quantify this agreement, we compute the Jensen-Shannon divergence (JSD) \cite{JS_div} between the original and reweighted marginalized posteriors for all 70 injections from large-$\Xeff$ population.
    For $\Xeff$, the JSD in bits has a median value of $1.7\times10^{-3}$ ranging from $1.6\times10^{-4}$ to $1.4\times10^{-2}$ 
    For $\Xp$, the median JSD in bits is $6.5\times10^{-3}$ ranging from $2.1\times10^{-3}$ to $2.3\times10^{-2}$. Each JSD is computed using approximately 5,000 posterior samples. These small divergences indicate that the inferred $(\Xeff,\Xp)$ posteriors, and hence the corresponding marginalized likelihood in the $(\Xeff,\Xp)$ parameterization, are largely insensitive to the choice of residual-spin completion.

    \subsection{Sensitivity to Monte Carlo sampling}
    
    To quantify the effect of finite Monte Carlo sampling, we examine the effective sample size, $N_\mathrm{eff}$.

    \begin{figure}[t]
    \centering
    \includegraphics[width=0.5\textwidth]{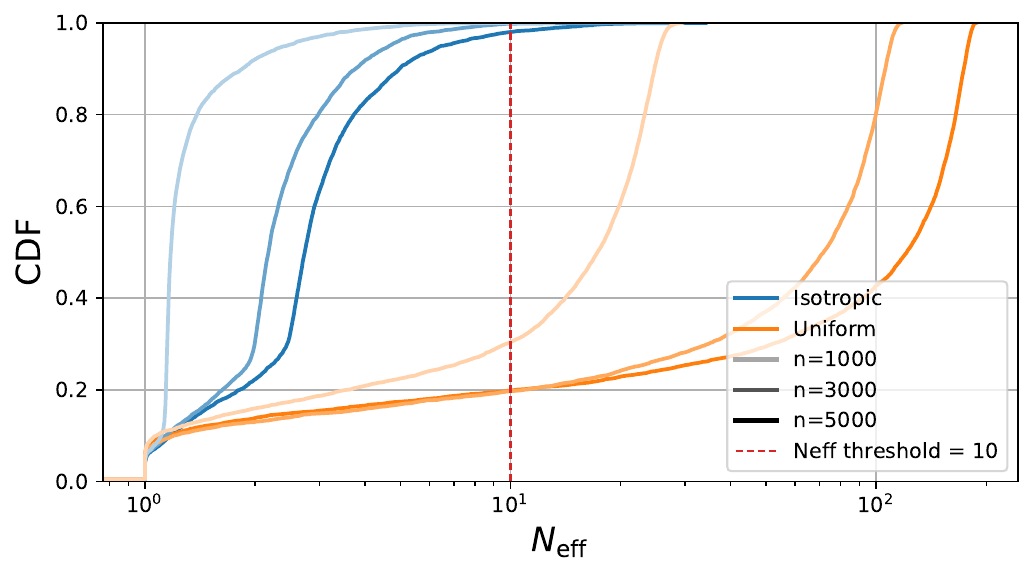}
    \caption{
    Cumulative distribution of the minimum $N_\mathrm{eff}$ across all events for hyperparameter samples drawn from the prior. Here, $N_\mathrm{eff}$ is given by the minimum effective sample size over all events. The cumulative probability at a given value of $N_\mathrm{eff}$ represents the fraction of hyperparameter samples with $N_\mathrm{eff}$ less than that value. The color opacity indicates the number of posterior samples for each event. The $n=5000$ case is the same as that used in our baseline analysis. The red vertical dashed line indicates the adopted $N_\mathrm{eff}$ threshold.}
    \label{total_neff_cdf_from_hyper_prior}
    \end{figure}

    Fig.~\ref{total_neff_cdf_from_hyper_prior} shows the cumulative distribution of the minimum $N_\mathrm{eff}$ across all events for hyperparameter samples drawn from the prior, while fixing all the other hyperparameters to their injection values listed in Table~\ref{population_hyperparameters}. We use the posterior samples from large-$\Xeff$ population analysis under the two event-level prior choices. The fraction of hyperparameter samples rejected by the $N_\mathrm{eff}$ threshold is 95\% for the isotropic spin prior, compared to only 22\% for the $\Xeff$-$\Xp$ uniform prior with $n=5000$, corresponding to the baseline analysis.
    As the number of posterior samples increases, the discard rate for the $\Xeff$-$\Xp$ uniform prior saturates at approximately 20\%. In contrast, the isotropic spin prior exhibits little further reduction in the discard rate, indicating that it remains inefficient for this region of parameter space.
    This demonstrates that the uniform prior substantially reduces Monte Carlo degradation over the hyperparameter space explored during population inference.

    Fig.~\ref{event_only_discard_corner} shows the regions rejected by the $N_\mathrm{eff}$ criterion in the large-$\Xeff$ population analysis. We focus only on the hyperparameters $\mu_\mathrm{p}$ and $\sigma_\mathrm{p}$, while fixing all other hyperparameters to their injection values. The colored regions correspond to the rejection regions for the isotropic spin and $\Xeff$-$\Xp$ uniform analyses. For the isotropic prior, regions with $\sigma_\mathrm{p} \lesssim 0.4$ are effectively rejected, biasing the inferred population distribution (Fig.~\ref{pop_iso_vs_uni_spin}). In contrast, the $\Xeff$-$\Xp$ uniform prior does not reject the injection value and substantially reduces the rejected region.

    \begin{figure}[t]
    \centering
    \includegraphics[width=0.4\textwidth]{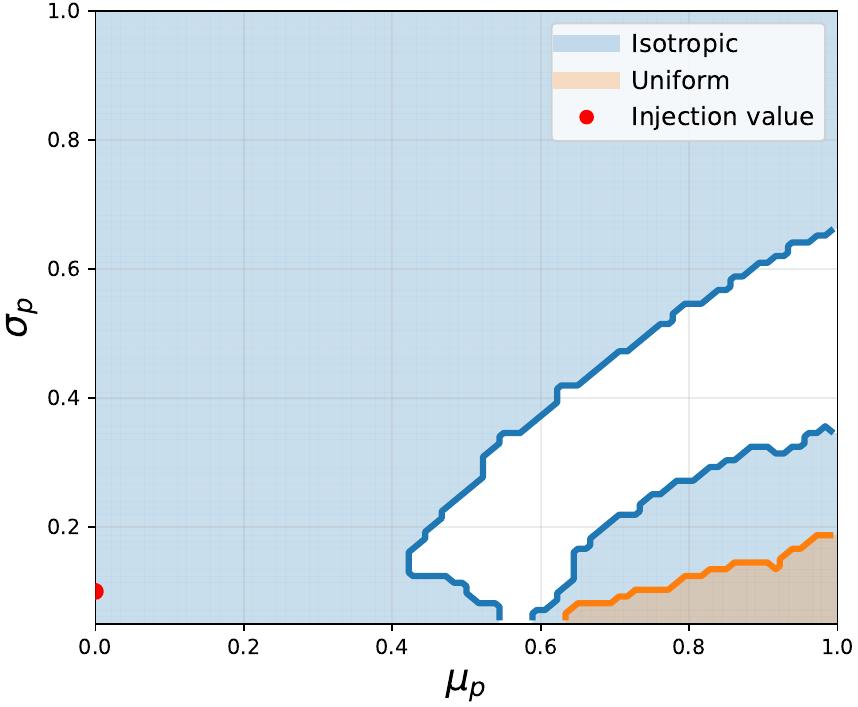}
    \caption{$N_{\mathrm{eff}}$ discarded region in the hyperparameters $\mu_\mathrm{p}$ and $\sigma_\mathrm{p}$ space in large-$\Xeff$ population analysis. These hyperparameters are drawn from the hyperparameter-prior, while all the other hyperparameters are fixed to their injection values. For each prior draw, we compute the $N_{\mathrm{eff}}$, and mark the draw as rejected when $N_{\mathrm{eff}}<10$. The rejection based on the $N_\mathrm{eff}$ criterion is not applied when computing the selection effect, since the selection function is identical for both event-level prior choices. Blue and orange correspond to the isotropic and uniform prior analyses, respectively. The red dot indicates the injection value. }
    \label{event_only_discard_corner}
    \end{figure}



    \end{widetext}

\bibliography{ref}%

\end{document}